\documentclass[11pt]{article}
\usepackage{amssymb}

\usepackage{graphicx}
\usepackage{amsmath}
\usepackage{makeidx}
\usepackage{graphicx}
\usepackage{graphicx}
\usepackage{indentfirst}

\newcounter{resultnum}[section]

\newcounter{conclusionnum}[section]

\newcounter{conditionnum}[section]

\newcounter{conjecturenum}[section]

\newcounter{examplenum}[section]

\newcounter{exercisenum}[section]

\newcounter{lemmanum}[section]

\newcounter{notationnum}[section]

\newcounter{theoremnum}[section]

\newcounter{definitionnum}[section]

\newcounter{corollarynum}[section]

\newcounter{remarknum}[section]

\newcounter{propositionnum}[section]

\newcounter{acknowledgementnum}[section]

\newcounter{algorithmnum}[section]

\newcounter{axiomnum}[section]

\newcounter{casenum}[section]

\newcounter{claimnum}[section]

\newcounter{summarynum}[section]

\newcounter{problemnum}[section]

\begin{document}

\title{Nonholonomic Ricci Flows:\\  Exact Solutions and Gravity }
\date{February 17, 2009}

\author{ Sergiu I. Vacaru
\thanks{
the affiliation for Fields Institute is for a former visiting position;  \newline
Sergiu.Vacaru@gmail.com;\   http://www.scribd.com/people/view/1455460-sergiu
 } \\
{\quad} \\
{\small {\textsl{Faculty of Mathematics, University "Al. I. Cuza" Ia\c si},}
}\\
{\small {\textsl{\ 700506, Ia\c si, Romania}} }\\
{\small and}\\
{\small \textsl{The Fields Institute for Research in Mathematical Science}}
\\
{\small \textsl{222 College Street, 2d Floor, Toronto \ M5T 3J1, Canada}} }

\maketitle

\begin{abstract}
In a number of physically important cases, the nonholonomically (nonintegrable) constrained Ricci flows can be modelled by exact solutions of Einstein equations with nonhomogeneous (anisotropic) cosmological constants. We develop two geometric methods for constructing such solutions: The first approach applies the formalism of nonholonomic frame deformations
when the gravitational evolution and field equations transform into systems of nonlinear partial differential equations which can be integrated in general form. The second approach develops a general scheme when one (two) parameter families of exact solutions are defined by any source-free solutions of Einstein's equations with one (two) Killing vector field(s). A
successive iteration procedure results in a class of solutions characterized by an infinite number of parameters for a non--Abelian group involving arbitrary functions on one variable. We also consider nonlinear superpositions of some mentioned classes of solutions in order to construct
more general integral varieties of the Ricci flow and Einstein equations depending on infinite number of parameters and three/ four coordinates on four/ five dimensional (semi) Riemannian spaces.

\vskip0.2cm

\textbf{Keywords:}\newline
Ricci flows, exact solutions, nonholonomic frames, nonlinear connections.

\vskip3pt

PACS Classification:

04.20.Jb, 04.30.Nk, 04.50.+h, 04.90.+e, 02.30.Jk, 02.40.-k

\vskip2pt

2000 AMS Subject Classification:

53A99, 53B40, 53C12, 53C44, 83C15, 83C20, 83C99, 83E99
\end{abstract}

\tableofcontents

\section{ Introduction}

In recent years, much work has been done on Ricci flow theory and
fundamental problems in mathematics \cite{4ham1,4ham2,4per1} (see \cite%
{4caozhu,4kleiner,4rbook} for reviews and references therein). In this
context, a number of possible applications in modern gravity and
mathematical physics were proposed, for instance, for low dimensional
systems and gravity \cite{4ni,4cv,4gk,4osw} and black holes and cosmology %
\cite{4hw,4bop}. Such special cases were investigated following certain low
dimensional or approximative solutions of the evolution equations.

There were also examined possible connections between Ricci flows, solitonic
configurations and Einstein spaces \cite{4nhrf01,4nhrf02,4nhrf03}. In our
works, we tackled the problem of constructing exact solutions in Ricci flow
and gravity theories in a new way. Working with general (pseudo) Riemannian
spaces and moving frames, we applied certain methods from the geometry of
Finsler--Lagrange spaces and nonholonomic manifolds provided with nonlinear
connection structure (N--connection) \cite{4ma,4bejf,4vesnc,4vsgg}.

Prescribing on a manifold some preferred systems of reference and
symmetries, it is equivalent to define some nonintegrable (nonholonomic,
equivalently, anholonomic) distributions with associated N--connections.
From this point of view, of the geometry of so--called nonholonomic
manifolds, it is possible to elaborate a unified formalism for locally
fibred manifolds and vector/tangent bundles when the geometric constructions
are adapted to the N--connection structure. We can consider different
classes of metric and N--connection ansatz and model, for instance, a
Finsler, or Lagrange, geometry in a (semi) Riemannian (in particular,
Einstein) space. Inversely, we can define some effective Lagrangian, or
Finsler like, fundamental functions for lifts of geometric objects for a
theory of gravity to tangent bundles in order to elaborate a geometric
mechanics model for such gravitational and/or gauge field interactions, see
examples and details in Refs. \cite{4vsgg,4vhep,4vt,4vp,4vs}. It was also
proved that constraining some classes of Ricci flows of (semi) Riemannian
metrics we can model Finsler like geometries and, inversely, we can
transform Finsler--Lagrange metrics and connections into Riemannian, or
Riemann--Cartan ones \cite{4nhrf01,4nhrf02,4entrnf}.

The most important idea in constructing exact solutions by geometric methods
is that we can consider such nonholonomic deformations of the frame and
connection structures when the Cartan structure equations, Ricci flow and/or
Einstein equations transform into systems of partial differential equations
which can be integrated in general form, or one can be derived certain
bi--Hamilton and solitonic equations with corresponding hierarchies and
conservation laws, see \cite{4vsgg,4nhrf03} and references therein.

The first examples of physically valuable exact solutions of nonholonomic
Ricci flow evolution equations and gravitational field equations were
constructed following the so--called anholonomic frame method \cite%
{4vrf,4vv1,4vv2}. We analyzed two general classes of solutions of evolution
equations on time like and/or extra dimension coordinate (having certain
nontrivial limits to exact solutions in gravity theories): The first class
was elaborated for solitonic and pp--wave nonholonomic configurations. The
second class was connected to a study of nonholonomic Ricci flow evolutions
of three and four dimensional (in brief, 3D and 4D) Taub--NUT metrics.
Following those constructions and further geometric developments in Refs. %
\cite{4nhrf01,4nhrf02}, we concluded that a number of important for physical
considerations solutions of Ricci flow equations can be defined by
nonholonomically generalized Einstein spaces with effective cosmological
constant running on evolution parameter, or (for more general and/or
normalized evolution flows) by 'nonhomogeneous' (locally anisotropic)
cosmological constants.

This is the forth paper in a series of works on nonholonomic Ricci flows
modelled by nonintegrable constraints on the frame structure and evolution
of metrics \cite{4nhrf01,4nhrf02,4nhrf03}. It is devoted to geometric
methods of constructing generic off--diagonal exact solutions in gravity and
Ricci flow theory.\footnote{%
We shall follow the conventions from the first two partner works in the
series; the reader is recommended to study them in advance.} The goal is to
elaborate a general scheme when starting with certain classes of metrics,
frames and connections new types of exact solutions are constructed
following some methods from nonholonomic spaces geometry \cite%
{4vhep,4vt,4vp,4vs,4vesnc} and certain group ideas \cite{4geroch1,4geroch2}.
The approach to generating vacuum Einstein metrics by parametric
nonholonomic transforms was recently formulated in Ref. \cite{4vpnhf} (this
article proposes a "Ricci flow development" of sections 2 and 3 in that
paper). Such results seem to have applications in modern gravity and
nonlinear physics: In the fifth partner paper \cite{4nhrf05}, we show how
nonholonomic Ricci flow evolution scenaria of physically valuable metrics
can be modelled by parametric deformations of solitonic pp--waves and
Schwarzschild solutions.

One should be noted that even there were found a large number of exact
solutions in different models of gravity theory \cite%
{4kramer,4bic,4vsgg,4vhep,4vt,4vp,4vs}, and in certain cases in the Ricci
flow theory \cite{4ni,4cv,4gk,4osw,4vrf,4vv1,4vv2,4nhrf03}, one has been
elaborated only a few general methods for generating new physical solutions
from a given metric describing a real physical situation. For quantum
fields, there were formulated some approximated approaches when (for
instance, by using Feynman diagrams, the formalism of Green's functions, or
quantum integrals) the solutions are constructed to represent a linear or
nonlinear prescribed physical situation. Perhaps it is unlikely that similar
computation techniques can be elaborated in general form in gravity theories
and related evolution equations. Nevertheless, certain new possibilities
seem to be opened after formulation of the anholonomic frame method with
parametric deformations for the Ricci flow theory. Although many of the
solutions resulting from such methods have no obvious physical
interpretation, one can be formulated some criteria selecting explicit
classes of solutions with prescribed symmetries and physical properties.

The paper has the following structure:\ In section 2, we outline some
results on nonholonomic manifolds and Ricci flows. Section 3 is devoted to
the anholonomic frame method for constructing exact solutions of Einstein
and Ricci flow equations. There are analyzed the conditions when such
solutions define four and five dimensional foliations related to Einstein
spaces and Ricci flows for the canonical distinguished connection and the
Levi Civita connection. In section 4, we consider how various classes of
metrics can be subjected to nonholonomic deformations and multi--parametric
transforms (with Killing symmetries) resulting in new classes of solutions
of the Einstein/ Ricci flow equations. We consider different ansatz for
metrics and two examples with multi--parametric families of Einstein spaces
and related Ricci flow evolution models. The results are discussed in
section 5. The reader is suggested to see Appendices before starting  the
main part of the paper:\ Appendix A outlines the geometry of nonlinear
connections and the anholonomic frame method of constructing exact
solutions. Appendix B summarizes some results on the parametric (Geroch)
transforms of vacuum Einstein equations.

\vskip3pt \textbf{Notation remarks:\ } It is convenient to use in parallel
two types of denotations for the geometric objects subjected to Ricci flows
by introducing \ ''left--up'' labels like $~^{\chi }\gamma =\gamma (...,\chi
).$ Different left--up labels will be also considered for some classes of
metrics defining Einstein spaces, vacuum solutions and so on. We shall also
write ''boldface'' symbols for geometric objects and spaces adapted to a
noholonomic (N--connection) structure, for instance, $\mathbf{V,E},...$ A
nonholonomic distribution with associated N--connection structure splits the
manifolds into conventional horizontal (h) and vertical (v) subspaces. The
geometric objects, for instance, a vector $\mathbf{X}$ can be written in
abstract form $\mathbf{X}=(hX,vX),$ or in coefficient forms, $\mathbf{X}%
^{\alpha }=(X^{i},X^{a})=(X^{\underline{i}},X^{\underline{a}}),$
equivalently decomposed with respect to a general nonholonomic frame $%
\mathbf{e}_{\alpha }=(e_{i},e_{a})$ or coordinate frame $\partial _{%
\underline{\alpha }}=(\partial _{\underline{i}},\partial _{\underline{a}})$
for local\ h- and v--coordinates $u=(x,y),$ or $u^{\alpha }=(x^{i},y^{a})$
when $\partial _{\underline{\alpha }}=\partial /\partial u^{\underline{%
\alpha }}$ and $\partial _{\underline{i}}=\partial /\partial x^{\underline{i}%
},$ $\partial _{\underline{a}}=\partial /\partial x^{\underline{a}}.$ The
h--indices $i,j,k,...=1,2,...n$ will be used for nonholonomic vector
components and the v--indices $a,b,c...=n+1,n+2,...n+m$ will be used for
holonomic vector components. Greek indices of type $\alpha ,\beta ,...$ will
be used as cumulative ones. We shall omit labels, indices and parametric/
coordinate dependencies if it does not result in ambiguities.

\section{Preliminaries}

In this section we present some results on nonholonomic manifolds and Ricci
flows \cite{4nhrf01,4nhrf02} selected with the aim to outline a new
geometric method of constructing exact solutions. The anholonomic frame
method and the geometry of nonlinear connections (N--connections) are
considered, in brief, in Appendix \ref{4sah}. The ideas on generating new
solutions from one/ two Killing vacuum Einstein spacetimes \cite%
{4geroch1,4geroch2} are summarized in Appendix \ref{4apkv}.

\subsection{Nonholonomic (pseudo) Riemannian spaces}

We consider a spacetime as a (necessary smooth class) manifold $V$ of
dimension $n+m,$ when $n\geq 2$ and $m\geq 1$ (a splitting of dimensions
being defined by a N--connection structure, see (\ref{4distr})). Such
manifolds (equivalently, spaces) are provided with a metric, $g=g_{\alpha
\beta }e^{\alpha }\otimes e^{\alpha },$ of any (pseudo) Euclidean signature
and a linear connection $D=\{\Gamma _{\ \beta \gamma }^{\alpha }e^{\beta }\}$
satisfying the metric compatibility condition $Dg=0.$\footnote{%
in this work, the Einstein's summation rule on repeating ''upper--lower''
indices will be applied if the contrary will not be stated} The components
of geometrical objects, for instance, $g_{\alpha \beta }$ and $\Gamma _{\
\beta \gamma }^{\alpha },$ are defined with respect to a local base (frame) $%
e_{\alpha }$ and its dual base (co--base, or co--frame) $e^{\alpha }$ for
which $e_{\alpha }\rfloor \ e^{\beta }=\delta _{\alpha }^{\beta },$ where ''$%
\rfloor "$ denotes the interior product induced by $g$ and $\delta _{\alpha
}^{\beta }$ is the Kronecker symbol. For a local system of coordinates $%
u^{\alpha }=(x^{i},y^{a})$ on $V$ (in brief, $u=(x,y)),$ we write
respectively
\begin{equation*}
e_{\alpha }=(e_{i}=\partial _{i}=\frac{\partial }{\partial x^{i}}%
,e_{a}=\partial _{a}=\frac{\partial }{\partial y^{a}})\mbox{ and }e^{\beta
}=(e^{j}=dx^{j},e^{b}=dy^{b}),
\end{equation*}%
for $e_{\alpha }\rfloor e^{\tau }=\delta _{\alpha }^{\tau };$ the indices
run correspondingly values of type:\ $i,j,...=1,2,...,n$ and $%
a,b,...=n+1,n+2,....,n+m$ for any conventional splitting $\alpha
=(i,a),\beta =(j,b),...$

Any local (vector) basis $e_{\alpha }$ and dual basis $e^{\beta}$ can be
decomposed with respect to other local bases $e_{\alpha ^{\prime }}$ and $%
e^{\beta ^{\prime }}$ by considering frame transforms,%
\begin{equation}
e_{\alpha }=A_{\alpha }^{\ \alpha ^{\prime }}(u)e_{\alpha ^{\prime }}%
\mbox{
and }e^{\beta }=A_{\ \beta ^{\prime }}^{\beta \ }(u)e^{\beta ^{\prime }},
\label{4ft}
\end{equation}%
where the matrix $A_{\ \beta ^{\prime }}^{\beta \ }$ is the inverse to $%
A_{\alpha }^{\ \alpha ^{\prime }}.$ It should be noted that an arbitrary
basis $e_{\alpha }$ is nonholonomic (equivalently, anholonomic) because, in
general, it satisfies certain anholonomy conditions
\begin{equation}
e_{\alpha }e_{\beta }-e_{\beta }e_{\alpha }=W_{\alpha \beta }^{\gamma }\
e_{\gamma }  \label{4anhr}
\end{equation}%
with nontrivial anholonomy coefficients $W_{\alpha \beta }^{\gamma }(u).$
For $W_{\alpha \beta }^{\gamma }=0,$ we get holonomic frames: for instance,
if we fix a local coordinate basis, $e_{\alpha }=\partial _{\alpha }.$

Denoting the covariant derivative along a vector field $X=X^{\alpha
}e_{\alpha }$ as $D_{X}=X\rfloor D,$ we can define the torsion
\begin{equation}
\mathcal{T}(X,Y)\doteqdot D_{X}Y-D_{Y}X-[X,Y],  \label{4tors}
\end{equation}%
and the curvature
\begin{equation}
\mathcal{R}(X,Y)Z\doteqdot D_{X}D_{Y}Z-D_{Y}D_{X}Z-D_{[X,Y]}Z,  \label{4curv}
\end{equation}%
tensors of connection $D,$ where we use ''by definition'' symbol ''$%
\doteqdot $'' and $[X,Y]\doteqdot XY-YX.$ The components $\mathcal{T}=\{T_{\
\beta \gamma }^{\alpha }\}$ and $\mathcal{R}=\{R_{\ \beta \gamma \tau
}^{\alpha }\}$ are computed by introducing $X\rightarrow e_{\alpha
},Y\rightarrow e_{\beta },Z\rightarrow e_{\gamma }$ into respective formulas
(\ref{4tors}) and (\ref{4curv}).

The Ricci tensor is constructed $Ric(D)=\{R_{\beta \gamma }\doteqdot R_{\
\beta \gamma \alpha }^{\alpha }\}.$ The scalar curvature $R$ is by
definition the contraction with $g^{\alpha \beta }$ (being the inverse to
the matrix $g_{\alpha \beta }),$ $R\doteqdot g^{\alpha \beta }R_{\alpha
\beta },$ and the Einstein tensor is $\mathcal{E}=\{E_{\alpha \beta
}\doteqdot R_{\alpha \beta }-\frac{1}{2}g_{\alpha \beta }R\}.$ The vacuum
(source--free) Einstein equations are
\begin{equation}
\mathcal{E}=\{E_{\alpha \beta }=R_{\alpha \beta }\}=0.  \label{4enstet}
\end{equation}

In general relativity theory, one chooses a connection $D=\nabla $ which is
uniquely defined by the coefficients of a metric, $g_{\alpha \beta },$
following the conditions of metric compatibility, $\nabla g=0,$ and of zero
torsion, $_{\shortmid }\mathcal{T}=0.$ This is the so--called Levi Civita
connection $\ _{\shortmid }D=\nabla .$ We shall respectively label its
curvature tensor, Ricci tensor, scalar curvature and Einstein tensor in the
form $\ _{\shortmid }\mathcal{R}=\{\ _{\shortmid }R_{\ \beta \gamma \tau
}^{\alpha }\},$ $\ _{\shortmid }Ric(\nabla )=\{\ _{\shortmid }R_{\alpha
\beta }\doteqdot \ _{\shortmid }R_{\ \beta \gamma \alpha }^{\alpha }\},$ $\
_{\shortmid }R\doteqdot g^{\alpha \beta }\ _{\shortmid }R_{\alpha \beta }$
and $\ _{\shortmid }\mathcal{E}=\{\ _{\shortmid }E_{\alpha \beta }\}.$

Modern gravity theories consider extra dimensions and connections with
nontrivial torsion. For instance, in string gravity \cite{4string1,4string2}%
, the torsion coefficients are induced by the so--called anti--symmetric $H$%
--fields and contain additional information about additional interactions in
low--energy string limit. A more special class of gravity interactions are
those with effective torsion when such fields are induced as a nonholonomic
frame effect in a unique form (\ref{4tors}) by prescribing a nonholonomic
distribution (\ref{4distr}), defining a nonlinear connection structure,
N--connection, on a (pseudo) Riemannian manifold \ $\mathbf{V}$ enabled with
a metric structure (\ref{4ansatz}). Such spaces with local fibred structure
are called nonholonomic (in more special cases, when the nonholonomy is
defined by a N--connection structures, the manifolds are called
N--anholonomic) \cite{4bejf,4vsgg}. The N--anholonomic spaces can be
described in equivalent form by two linear connections $\nabla $ (\ref%
{4candcon}) and $\widehat{\mathbf{D}}$ $\ $(\ref{4lccon}), both metric
compatible and completely stated by a metric (\ref{4dmetr}), equivalently (%
\ref{4metr}). \footnote{%
In this work, we shall use only $\nabla $ and the canonical d--connection $%
\widehat{\mathbf{D}}$ and, for simplicity, we shall omit ''hat'' writing $%
\mathbf{D}$ if that will not result in ambiguities.} As a matter of
principle, the general relativity theory can be formulated in terms of both
connections, $\nabla $ and $\widehat{\mathbf{D}};$ the last variant being
with nonholonomic constraints on geometrical objects. One must be emphasized
that the standard approach follows the formulation of gravitational field
equations just for the Einstein tensor $\ _{\shortmid }\mathcal{E}$ for $%
\nabla $ which, in general, is different from the Einstein tensor $\widehat{%
\mathbf{E}}$ for $\widehat{\mathbf{D}}.$

A surprising thing found in our works is that for certain classes of generic
off--diagonal metric ansatz (\ref{4ansatz}) it is possible to construct
exact solutions in general form by using the connection $\widehat{\mathbf{D}}
$ but not the connection $\nabla .$ Here we note that having defined certain
integral varieties for a first class of linear connections we can impose
some additional constraints and generate solutions for a class of Levi
Civita connections, for instance, in the Einstein and  string, or Finsler
like, generalizations of gravity. Following a geometric N--adapted formalism
(the so--called anholonomic frame method), such solutions were constructed
and studied in effective noncommutative gravity \cite{4vesnc}, various
locally anisotropic (Finsler like and more general ones) extensions of the
Einstein and Kaluza--Klein theory, in string an brane gravity \cite%
{4vhep,4vt,4vs} and for Lagrange--Fedosov manifolds \cite{4esv}, see a
summary in \cite{4vsgg}.

The anholonomic frame method also allows us to construct exact solutions in
general relativity: One defines a more general class of solutions for $%
\widehat{\mathbf{D}}$ and then imposes certain subclasses of nonholonomic
constraints when such solutions solve the four dimensional Einstein
equations for $\nabla .$ Here we note that by nonholonomic deformations we
were able to study nonholonomic Ricci flows of certain classes of physically
valuable exact solutions like solitonic pp--waves \cite{4vrf} and Taub NUT
spaces \cite{4vv1,4vv2}. In this work, we develop the approach by applying
new group methods.

Certain nontrivial limits to the vacuum Einstein gravity can be selected if
we impose on the nonholonomic structure such constraints when
\begin{equation}
\mathbf{E}=\ _{\shortmid }\mathcal{E}  \label{4cond1}
\end{equation}%
even, in general, $\mathbf{D}\neq \nabla .$ We shall consider such
conditions when $\mathbf{D}$ and $\nabla $ have the same components with
respect to certain preferred bases and the equality (\ref{4cond1}) can be
satisfied for some very general classes of metric ansatz.\footnote{%
We emphasize that different linear connections may be subjected to different
rules of frame and coordinate transforms. It should be noted here that
tensors and nonlinear and linear connections transform in different ways
under frame and coordinate changing on manifolds with locally fibred
structures, see detailed discussions in \cite{4ma,4vsgg}.}

We shall use left--up labels ''$\circ $'' or ''$^{\lambda }$'' for a metric,
\begin{equation*}
\ ^{\circ }g=\ ^{\circ }g_{\alpha \beta }\ e^{\alpha }\otimes e^{\beta }%
\mbox{ or }\ ^{\lambda }g=\ ^{\lambda }g_{\alpha \beta }\ e^{\alpha }\otimes
e^{\beta }
\end{equation*}%
being (correspondingly) a solution of the vacuum Einstein (or with
cosmological constant) equations $\mathcal{E}=0$ (\ref{4enstet}) or of the
Einstein equations with a cosmological constant $\lambda ,$ $R_{\alpha \beta
}=\lambda g_{\alpha \beta },$ for a linear connection $D$ with possible
torsion $\mathcal{T}\neq 0.$ In order to emphasize that a metric is a
solution of the vacuum Einstein equations, in any dimension $n+m\geq 3,$ for
the Levi Civita connection $\nabla $, we shall write
\begin{equation*}
\ _{\shortmid }^{\circ }g=\ _{\shortmid }^{\circ }g_{\alpha \beta }\
e^{\alpha }\otimes e^{\beta }\mbox{ or }\ _{\shortmid }^{\lambda }g=\
_{\shortmid }^{\lambda }g_{\alpha \beta }\ e^{\alpha }\otimes e^{\beta },
\end{equation*}%
where the left--low label $"_{\shortmid }"$ will distinguish the geometric
objects for the Ricci flat space defined by a Levi Civita connection $\nabla
.$

Finally, in this section, we note that we shall use ''boldface'' symbols,
for instance, if $\ ^{\lambda }\mathbf{g}=\ ^{\lambda }\mathbf{g}_{\alpha
\beta }\ \mathbf{e}^{\alpha }\otimes \mathbf{e}^{\beta }$ defines a
nonholonomic Einstein space as a solution of%
\begin{equation}
\mathbf{R}_{\alpha \beta }=\lambda \mathbf{g}_{\alpha \beta }  \label{4ep1b}
\end{equation}
for the canonical d--connection $\mathbf{D.}$

\subsection{Evolution equations for nonholonomic Ricci flows}

The normalized (holonomic) Ricci flows \cite{4per1,4caozhu,4kleiner,4rbook}
for a family of metrics $g_{\underline{\alpha }\underline{\beta }}(\chi )=g_{%
\underline{\alpha }\underline{\beta }}(u^{\underline{\nu }},\chi ),$
parametrized by a real parameter $\chi ,$ with respect to the coordinate
base $\partial _{\underline{\alpha }}=\partial /\partial u^{\underline{%
\alpha }},$ are described by the equations
\begin{equation}
\frac{\partial }{\partial \chi }g_{\underline{\alpha }\underline{\beta }%
}=-2\ _{\shortmid }R_{\underline{\alpha }\underline{\beta }}+\frac{2r}{5}g_{%
\underline{\alpha }\underline{\beta }},  \label{4feq}
\end{equation}%
where the normalizing factor $r=\int \ _{\shortmid }RdV/dV$ is introduced in
order to preserve the volume $V.$ \footnote{%
The Ricci flow evolution equations were introduced by R. Hamilton \cite%
{4ham1}, as evolution equations
\begin{equation*}
\frac{\partial \underline{g}_{\alpha \beta }(\chi )}{\partial \chi }=-2\
_{\shortmid }\underline{R}_{\alpha \beta }(\chi ),
\end{equation*}%
for a set of Riemannian metrics $\underline{g}_{\alpha \beta }(\chi )$ and
corresponding Ricci tensors $\ _{\shortmid }\underline{R}_{\alpha \beta
}(\chi )$ parametrized by a real $\chi $ (we shall underline symbols or
indices in order to emphasize that certain geometric objects/ equations are
given with the components defined with respect to a coordinate basis). For
our further purposes, on generalized Riemann--Finsler spaces, it is
convenient to use a different system of denotations than those considered by
R. Hamilton or Grisha Perelman on holonomic Riemannian spaces.}

For N--anholonomic Ricci flows, the coefficients $g_{\underline{\alpha }%
\underline{\beta }}$ are parametrized in the form (\ref{4ansatz}), see
proofs and discussion in Refs. \cite{4nhrf01,4vrf,4vv1,4vv2}. With respect
to N--adapted frames (\ref{4dder}) and (\ref{4ddif}), the Ricci flow
equations (\ref{4feq}), redefined for $\nabla \rightarrow \widehat{\mathbf{D}%
}$ and, respectively, $_{\shortmid }R_{\alpha \beta }\rightarrow \widehat{%
\mathbf{R}}_{\alpha \beta }$ are
\begin{eqnarray}
\frac{\partial }{\partial \chi }g_{ij} &=&2\left[ N_{i}^{a}N_{j}^{b}\ \left(
\underline{\widehat{R}}_{ab}-\lambda g_{ab}\right) -\underline{\widehat{R}}%
_{ij}+\lambda g_{ij}\right] -g_{cd}\frac{\partial }{\partial \chi }%
(N_{i}^{c}N_{j}^{d}),  \label{4e1} \\
\frac{\partial }{\partial \chi }g_{ab} &=&-2\ \left( \underline{\widehat{R}}%
_{ab}-\lambda g_{ab}\right) ,\   \label{4e2} \\
\ \widehat{R}_{ia} &=&0\mbox{ and }\ \widehat{R}_{ai}=0,  \label{4e3}
\end{eqnarray}%
where $\lambda =r/5$ the Ricci coefficients $\underline{\widehat{R}}_{ij}$
and $\underline{\widehat{R}}_{ab}$ are computed with respect to coordinate
coframes. The equations (\ref{4e3}) constrain the nonholonomic Ricci flows
to result in symmetric metrics.\footnote{%
In Refs. \cite{4nhrf01,4entrnf}, we discuss this problem related to the fact
that the tensor $\widehat{\mathbf{R}}_{\alpha \beta }$ is not symmetric
which results, in general, in Ricci flows of nonsymmetric metrics.}

Nonholonomic deformations of geometric objects (and related systems of
equations) on a N--anholonomic manifold $\mathbf{V}$ are defined for the
same metric structure $\mathbf{g}$ by a set of transforms of arbitrary
frames into N--adapted ones and of the Levi Civita connection $\nabla $ into
the canonical d--connection $\widehat{\mathbf{D}},$ locally parametrized in
the form
\begin{equation*}
\partial _{\underline{\alpha }}=(\partial _{\underline{i}},\partial _{%
\underline{a}})\rightarrow \mathbf{e}_{\alpha }=(\mathbf{e}_{i},e_{a});\ g_{%
\underline{\alpha }\underline{\beta }}\rightarrow \lbrack
g_{ij},g_{ab},N_{i}^{a}];\ _{\shortmid }\Gamma _{\ \alpha \beta }^{\gamma
}\rightarrow \widehat{\mathbf{\Gamma }}_{\ \alpha \beta }^{\gamma }.
\end{equation*}%
A rigorous proof for nonholonomic evolution equations is possible following
a N--adapted variational calculus for the Perelman's functionals presented
in Refs. \cite{4nhrf02}. For a five dimensional space with diagonal
d--metric ansatz (\ref{4dmetr}), when $g_{ij}=diag[\pm 1,g_{2},g_{3}]$ and $%
g_{ab}=diag[g_{4},g_{5}],$ we considered \cite{4vrf} the nonholonomic
evolution equations
\begin{eqnarray}
\frac{\partial }{\partial \chi }g_{ii} &=&-2\left[ \widehat{R}_{ii}-\lambda
g_{ii}\right] -g_{cc}\frac{\partial }{\partial \chi }(N_{i}^{c})^{2},
\label{4eq1} \\
\frac{\partial }{\partial \chi }g_{aa} &=&-2\ \left( \widehat{R}%
_{aa}-\lambda g_{aa}\right) ,\   \label{4eq2} \\
\ \widehat{R}_{\alpha \beta } &=&0\mbox{ for }\ \alpha \neq \beta ,
\label{4eq3}
\end{eqnarray}%
with the coefficients defined with respect to N--adapted frames (\ref{4dder}%
) and (\ref{4ddif}). This system can be transformed into a similar one, like
(\ref{4e1})--(\ref{4e3}), by nonholonomic deformations.

\section{Off--Diagonal Exact Solutions}

We consider a five dimensional (5D) manifold $\mathbf{V}$ of necessary
smooth class and conventional splitting of dimensions $\dim \mathbf{V=}$ $%
n+m $ for $n=3$ and $m=2.$ The local coordinates are labelled in the form $%
u^{\alpha }=(x^{i},y^{a})=(x^{1},x^{\widehat{i}},y^{4}=v,y^{5}),$ for $%
i=1,2,3$ and $\widehat{i}=2,3$ and $a,b,...=4,5.$ Any coordinates from a set
$u^{\alpha }$ can be a three dimensional (3D) space, time, or extra
dimension (5th) one. Ricci flows of geometric objects will be parametrized
by a real $\chi .$

\subsection{Off--diagonal ansatz for Einstein spa\-ces and Ricci flows}

The ansatz of type (\ref{4dmetr}) is parametrized in the form
\begin{eqnarray}
\mathbf{g} &=&g_{1}{dx^{1}}\otimes {dx^{1}}+g_{2}(x^{2},x^{3}){dx^{2}}%
\otimes {dx^{2}}+g_{3}\left( x^{2},x^{3}\right) {dx^{3}}\otimes {dx^{3}}
\notag \\
&&+h_{4}\left( x^{k},v\right) \ {\delta v}\otimes {\delta v}+h_{5}\left(
x^{k},v\right) \ {\delta y}\otimes {\delta y},  \notag \\
\delta v &=&dv+w_{i}\left( x^{k},v\right) dx^{i},\ \delta y=dy+n_{i}\left(
x^{k},v\right) dx^{i}  \label{4ans5d}
\end{eqnarray}%
with the coefficients defined by some necessary smooth class functions
\begin{eqnarray}
g_{1} &=&\pm 1,g_{2,3}=g_{2,3}(x^{2},x^{3}),h_{4,5}=h_{4,5}(x^{i},v),  \notag
\\
w_{i} &=&w_{i}(x^{i},v),n_{i}=n_{i}(x^{i},v).  \notag
\end{eqnarray}%
The off--diagonal terms of this metric, written with respect to the
coordinate dual frame $du^{\alpha }=(dx^{i},dy^{a}),$ can be redefined to
state a N--connection structure $\mathbf{N}=[N_{i}^{4}=w_{i}(x^{k},v),$$%
N_{i}^{5}=n_{i}(x^{k},v)]$ with a N--elongated co--frame (\ref{4ddif})
parametrized as
\begin{eqnarray}
e^{1} &=&dx^{1},\ e^{2}=dx^{2},\ e^{3}=dx^{3},  \notag \\
\mathbf{e}^{4} &=&\delta v=dv+w_{i}dx^{i},\ \mathbf{e}^{5}=\delta
y=dy+n_{i}dx^{i}.  \label{4ddif5}
\end{eqnarray}%
This coframe is dual to the local basis%
\begin{equation}
\mathbf{e}_{i}=\frac{\partial }{\partial x^{i}}-w_{i}\left( x^{k},v\right)
\frac{\partial }{\partial v}-n_{i}\left( x^{k},v\right) \frac{\partial }{%
\partial y^{5}},e_{4}=\frac{\partial }{\partial v},e_{5}=\frac{\partial }{%
\partial y^{5}}.  \label{4dder5}
\end{equation}%
We emphasize that the metric (\ref{4ans5d}) does not depend on variable $%
y^{5},$ i.e. it posses a Killing vector $e_{5}=\partial /\partial y^{5},$
and distinguishes the dependence on the so--called ''anisotropic'' variable $%
y^{4}=v.$

The above considered ansatz and formulas can be generalized in order to
model Ricci flows,
\begin{eqnarray}
~^{\chi }\mathbf{g} &=&g_{1}{dx^{1}}\otimes {dx^{1}}+g_{2}(x^{2},x^{3},\chi )%
{dx^{2}}\otimes {dx^{2}}+g_{3}\left( x^{2},x^{3},\chi \right) {dx^{3}}%
\otimes {dx^{3}}  \notag \\
&&+h_{4}\left( x^{k},v,\chi \right) \ ~^{\chi }{\delta v}\otimes ~^{\chi }{%
\delta v}+h_{5}\left( x^{k},v,\chi \right) \ ~^{\chi }{\delta y}\otimes
~^{\chi }{\delta y},  \notag \\
~^{\chi }\delta v &=&dv+w_{i}\left( x^{k},v,\chi \right) dx^{i},\ ~^{\chi
}\delta y=dy+n_{i}\left( x^{k},v,\chi \right) dx^{i}  \label{4ans5dr}
\end{eqnarray}%
with corresponding flows for N--adapted bases,
\begin{eqnarray*}
\mathbf{e}_{\alpha } &=&(\mathbf{e}_{i},e_{a})\rightarrow ~^{\chi }\mathbf{e}%
_{\alpha }=(~^{\chi }\mathbf{e}_{i},e_{a})=\mathbf{e}_{\alpha }(\chi )=(%
\mathbf{e}_{i}(\chi ),e_{a}), \\
\mathbf{e}^{\alpha } &=&(e^{i},\mathbf{e}^{a})\rightarrow ~^{\chi }\mathbf{e}%
^{\alpha }=(e^{i},~^{\chi }\mathbf{e}^{a})=\mathbf{e}^{\alpha }(\chi
)=(e^{i},\mathbf{e}^{a}(\chi ))
\end{eqnarray*}%
defined by $w_{i}\left( x^{k},v\right) \rightarrow w_{i}\left(
x^{k},v,\lambda \right), $ $n_{i}\left( x^{k},v\right) \rightarrow
n_{i}\left( x^{k},v,\lambda \right) $ in (\ref{4dder5}), (\ref{4ddif5}).

Computing the components of the Ricci and Einstein tensors for the metric (%
\ref{4ans5dr}) (see main formulas in Appendix and details on tensors
components' calculus in Refs. \cite{4vesnc,4vsgg}), one proves that the
corresponding family of Ricc tensors for the canonical d--connection with
respect to N--adapted frames are compatible with the sources (they can be
any matter fields, string corrections, Ricci flow parameter derivatives of
metric, ...)%
\begin{eqnarray}
\mathbf{\Upsilon }_{\beta }^{\alpha } &=&[\Upsilon _{1}^{1}=\Upsilon
_{2}+\Upsilon _{4},\Upsilon _{2}^{2}=\Upsilon _{2}(x^{2},x^{3},v,\chi
),\Upsilon _{3}^{3}=\Upsilon _{2}(x^{2},x^{3},v,\chi ),  \notag \\
\Upsilon _{4}^{4} &=&\Upsilon _{4}(x^{2},x^{3},\chi ),\Upsilon
_{5}^{5}=\Upsilon _{4}(x^{2},x^{3},\chi )]  \label{4sdiag}
\end{eqnarray}%
transform into this system of partial differential equations:
\begin{eqnarray}
R_{2}^{2} &=&R_{3}^{3}(\chi )  \label{4ep1a} \\
&=&\frac{1}{2g_{2}g_{3}}[\frac{g_{2}^{\bullet }g_{3}^{\bullet }}{2g_{2}}+%
\frac{(g_{3}^{\bullet })^{2}}{2g_{3}}-g_{3}^{\bullet \bullet }+\frac{%
g_{2}^{^{\prime }}g_{3}^{^{\prime }}}{2g_{3}}+\frac{(g_{2}^{^{\prime }})^{2}%
}{2g_{2}}-g_{2}^{^{\prime \prime }}]=-\Upsilon _{4}(x^{2},x^{3},\chi ),
\notag \\
S_{4}^{4} &=&S_{5}^{5}(\chi )=\frac{1}{2h_{4}h_{5}}\left[ h_{5}^{\ast
}\left( \ln \sqrt{|h_{4}h_{5}|}\right) ^{\ast }-h_{5}^{\ast \ast }\right]
=-\Upsilon _{2}(x^{2},x^{3},v,\chi ),  \label{4ep2a} \\
R_{4i} &=&-w_{i}(\chi )\frac{\beta (\chi )}{2h_{5}(\chi )}-\frac{\alpha
_{i}(\chi )}{2h_{5}(\chi )}=0,  \label{4ep3a} \\
R_{5i} &=&-\frac{h_{5}(\chi )}{2h_{4}(\chi )}\left[ n_{i}^{\ast \ast }(\chi
)+\gamma (\chi )n_{i}^{\ast }(\chi )\right] =0,  \label{4ep4a}
\end{eqnarray}%
where, for $h_{4,5}^{\ast }\neq 0,$%
\begin{eqnarray}
\alpha _{i}(\chi ) &=&h_{5}^{\ast }(\chi )\partial _{i}\phi (\chi ),\ \beta
(\chi )=h_{5}^{\ast }(\chi )\ \phi ^{\ast }(\chi ),\   \label{4coef} \\
\gamma (\chi ) &=&\frac{3h_{5}^{\ast }(\chi )}{2h_{5}(\chi )}-\frac{%
h_{4}^{\ast }(\chi )}{h_{4}(\chi )},~\phi (\chi )=\ln |\frac{h_{5}^{\ast
}(\chi )}{\sqrt{|h_{4}(\chi )h_{5}(\chi )|}}|,  \label{4coefa}
\end{eqnarray}%
when the necessary partial derivatives are written in the form $a^{\bullet
}=\partial a/\partial x^{2},$\ $a^{\prime }=\partial a/\partial x^{3},$\ $%
a^{\ast }=\partial a/\partial v.$ In the vacuum case, we must consider $%
\Upsilon _{2,4}=0.$ We note that we use a source of type (\ref{4sdiag}) in
order to show that the anholonomic frame method can be applied also for
non--vacuum configurations, for instance, when $\Upsilon _{2}=\lambda
_{2}=const$ and $\Upsilon _{4}=\lambda _{4}=const,$ defining local
anisotropies generated by an anisotropic cosmological constant, which in its
turn, can be induced by certain ansatz for the so--called $H$--field
(absolutely antisymmetric third rank tensor) in string theory \cite%
{4vesnc,4vsgg}. We note that the off--diagonal gravitational interactions
and Ricci flows can model locally anisotropic configurations even if $%
\lambda _{2}=\lambda _{4},$ or both values vanish.

Summarizing the results for an ansatz (\ref{4ans5d}) with arbitrary
signatures $\epsilon _{\alpha }=\left( \epsilon _{1},\epsilon _{2},\epsilon
_{3},\epsilon _{4},\epsilon _{5}\right) $ (where $\epsilon _{\alpha }=\pm 1)$
and $h_{4}^{\ast }\neq 0$ and $h_{5}^{\ast }\neq 0,$ for a fixed value of $%
\chi ,$ one proves, see details in \cite{4vesnc,4vsgg}, that any
off---diagonal metric
\begin{eqnarray}
\ ^{\circ }\mathbf{g} &=&\epsilon _{1}\ dx^{1}\otimes dx^{1}+\epsilon
_{2}g_{2}(x^{\widehat{i}})\ dx^{2}\otimes dx^{2}+\epsilon _{3}g_{3}(x^{%
\widehat{i}})\ dx^{3}\otimes dx^{3}  \notag \\
&&+\epsilon _{4}h_{0}^{2}(x^{i})\left[ f^{\ast }\left( x^{i},v\right) \right]
^{2}|\varsigma \left( x^{i},v\right) |\ \delta v\otimes \delta v  \notag \\
&&+\epsilon _{5}\left[ f\left( x^{i},v\right) -f_{0}(x^{i})\right] ^{2}\
\delta y^{5}\otimes \delta y^{5},  \notag \\
\delta v &=&dv+w_{k}\left( x^{i},v\right) dx^{k},\ \delta
y^{5}=dy^{5}+n_{k}\left( x^{i},v\right) dx^{k},  \label{4gensol1}
\end{eqnarray}%
with the coefficients being of necessary smooth class and the idices with
''hat'' running the values $\widehat{i},\widehat{j},...=2,3$, where\ \ $g_{%
\widehat{k}}\left( x^{\widehat{i}}\right) $ is a solution of the 2D equation
(\ref{4ep1a}) for a given source $\Upsilon _{4}\left( x^{\widehat{i}}\right)
,$%
\begin{equation*}
\varsigma \left( x^{i},v\right) =\varsigma _{\lbrack 0]}\left( x^{i}\right) -%
\frac{\epsilon _{4}}{8}h_{0}^{2}(x^{i})\int \Upsilon _{2}(x^{\widehat{k}%
},v)f^{\ast }\left( x^{i},v\right) \left[ f\left( x^{i},v\right)
-f_{0}(x^{i})\right] dv,
\end{equation*}%
and the N--connection coefficients $N_{i}^{4}=w_{i}(x^{k},v),$ $%
N_{i}^{5}=n_{i}(x^{k},v)$ are computed following the formulas
\begin{eqnarray}
w_{i}&=&-\frac{\partial _{i}\varsigma \left( x^{k},v\right) }{\varsigma
^{\ast }\left( x^{k},v\right) }  \label{4gensol1w} \\
n_{k}&=&n_{k[1]}\left( x^{i}\right) +n_{k[2]}\left( x^{i}\right) \int \frac{%
\left[ f^{\ast }\left( x^{i},v\right) \right] ^{2}}{\left[ f\left(
x^{i},v\right) -f_{0}(x^{i})\right] ^{3}}\varsigma \left( x^{i},v\right) dv,
\label{4gensol1n}
\end{eqnarray}%
define an exact solution of the system of Einstein equations (\ref{4ep1b}).
It should be emphasized that such solutions depend on arbitrary functions $%
f\left( x^{i},v\right),$ with $f^{\ast }\neq 0,$ $f_{0}(x^{i}),$ $%
h_{0}^{2}(x^{i})$, $\ \varsigma _{\lbrack 0]}\left( x^{i}\right) ,$ $%
n_{k[1]}\left( x^{i}\right), $ $\ n_{k[2]}\left( x^{i}\right)$ and $\Upsilon
_{2}(x^{\widehat{k}},v),\Upsilon _{4}\left( x^{\widehat{i}}\right) .$ Such
values for the corresponding signatures $\epsilon _{\alpha }=\pm 1$ have to
be stated by certain boundary conditions following some physical
considerations.\footnote{%
Our classes of solutions depending on integration functions are more general
than those for diagonal ansatz depending, for instance, on one radial like
variable like in the case of the Schwarzschild solution (when the Einstein
equations are reduced to an effective nonlinear ordinary differential
equation, ODE). In the case of ODE, the integral varieties depend on
integration constants to be defined from certain boundary/ asymptotic and
symmetry conditions, for instance, from the constraint that far away from
the horizon the Schwarzschild metric contains corrections from the Newton
potential. Because our ansatz (\ref{4ans5d}) transforms  (\ref{4ep1b}) in a
system of nonlinear partial differential equations transforms, the solutions
depend not on integration constants but on  integration functions.}

The ansatz of type (\ref{4ans5d}) with $h_{4}^{\ast }=0$ but $h_{5}^{\ast
}\neq 0$ (or, inversely, $h_{4}^{\ast }\neq 0$ but $h_{5}^{\ast }=0)$
consist more special cases and request a bit different methods for
constructing exact solutions. Nevertheless, such solutions are also generic
off--diagonal and they may be of substantial interest (the length of paper
does not allow us to include an analysis of such particular cases).

\subsection{Generalization of solutions for Ricci flows}

For families of solutions parametrized by $\chi ,$ we consider flows of the
generating functions, $g_{2}(x^{i},\chi),$ or $g_{3}(x^{i},\chi),$ and $%
f\left( x^{i},v,\chi \right) ,$ and various types of integration functions
and sources, for instance, $n_{k[1]}\left( x^{i},\chi \right) $ and $%
n_{k[2]}\left( x^{i},\chi \right) $ and $\Upsilon _{2}(x^{\widehat{k}%
},v,\chi ),$ respectively, in formulas (\ref{4gensol1w}) and (\ref{4gensol1n}%
). Let us analyze an example of exact solutions of equations (\ref{4eq1})--(%
\ref{4eq3}):

We search a class of solutions of with
\begin{eqnarray*}
g_{2} &=&\epsilon _{2}\varpi (x^{2},x^{3},\chi ),g_{3}=\epsilon _{3}\varpi
(x^{2},x^{3},\chi ), \\
h_{4} &=&h_{4}\left( x^{2},x^{3},v\right) ,h_{5}=h_{5}\left(
x^{2},x^{3},v\right) ,
\end{eqnarray*}%
for a family of ansatz (\ref{4ans5dr}) with any prescribed signatures $%
\epsilon _{\alpha }=\pm 1$ and non--negative functions $\varpi $ and $h.$
Following a tensor calculus, adapted to the N--connection, for the canonical
d--connection,\footnote{%
similar computations are given in \cite{4vesnc} and Chapter 10 of \cite%
{4vsgg}} we express the integral variety for a class of nonholonomic Ricci
flows as
\begin{eqnarray}
\epsilon _{2}(\ln \left| \varpi \right| )^{\bullet \bullet }+\epsilon
_{3}(\ln \left| \varpi \right| )^{^{\prime \prime }} &=&2\lambda
-h_{5}\partial _{\chi }\left( n_{2}\right) ^{2},  \label{4rfea1} \\
h_{4} &=&h\varsigma _{4}  \notag
\end{eqnarray}%
for%
\begin{eqnarray}
\varsigma _{4}(x^{2},x^{3},v) &=&\varsigma _{4[0]}(x^{2},x^{3})-\frac{%
\lambda }{4}\int \frac{hh_{5}}{h_{5}^{\ast }}dv  \notag \\
\sqrt{|h|} &=&h_{[0]}(x^{i})\left( \sqrt{|h_{5}\left( x^{2},x^{3},v\right) |}%
\right) ^{\ast }  \label{4rfea2}
\end{eqnarray}%
and, for $\varphi =-\ln \left| \sqrt{|h_{4}h_{5}|}/|h_{5}^{\ast }|\right|,$
\begin{eqnarray}
w_{1} &=&0,w_{2}=(\varphi ^{\ast })^{-1}\varphi ^{\bullet },w_{3}=(\varphi
^{\ast })^{-1}\varphi ^{\prime },  \label{4rfea3} \\
n_{1} &=&0,n_{2}=n_{3}=n_{[1]}(x^{2},x^{3},\chi )+n_{[2]}(x^{2},x^{3},\chi
)\int dv~h_{4}/\left( \sqrt{\left| h_{5}\right| }\right) ^{3},  \notag
\end{eqnarray}%
where the partial derivatives are denoted in the form $\varphi ^{\bullet
}=\partial \varphi /\partial x^{2},\varphi ^{^{\prime }}=\partial \varphi
/\partial x^{3},\varphi ^{\ast }=\partial \varphi /\partial v,\partial
_{\chi }=\partial /\partial x^{2},$ and arbitrary $h_{5}$ when $h_{5}^{\ast
}\neq 0.$ For $\lambda =0,$ we shall consider $\varsigma _{4[0]}=1$ and $%
h_{[0]}(x^{i})=const$ in order to solve the vacuum Einstein equations. There
is a class of solutions when
\begin{equation*}
h_{5}\int dv~h_{4}/\left( \sqrt{\left| h_{5}\right| }\right)
^{3}=C(x^{2},x^{3}),
\end{equation*}
for a function $C(x^{2},x^{3}).$  This is compatible with the condition (\ref%
{4rfea2}), and we can chose such configurations, for instance, with $%
n_{[1]}=0$ and any $n_{[2]}(x^{2},x^{3},\chi )$ and $\varpi
(x^{2},x^{3},\chi )$ solving the equation (\ref{4rfea1}).

Putting together (\ref{4rfea1})--(\ref{4rfea3}), we get a class of solutions
of the system (\ref{4eq1})--(\ref{4eq3}) (the equations being expressed
equivalently in the form (\ref{4ep1a})--(\ref{4ep4a})) for nonholonomomic
Ricci flows of metrics of type (\ref{4ans5dr}),
\begin{eqnarray}
~^{\chi }\mathbf{g} &=&\epsilon _{1}{dx^{1}}\otimes {dx^{1}}+\varpi
(x^{2},x^{3},\chi )\left[ \epsilon _{2}{dx^{2}}\otimes {dx^{2}}+\epsilon _{3}%
{dx^{3}}\otimes {dx^{3}}\right]  \notag \\
&&+h_{4}\left( x^{2},x^{3},v\right) \ {\delta v}\otimes {\delta v}%
+h_{5}\left( x^{2},x^{3},v\right) \ ~^{\chi }{\delta y}\otimes ~^{\chi }{%
\delta y},  \notag \\
\delta v &=&dv+w_{2}\left( x^{2},x^{3},v\right) dx^{2}+w_{3}\left(
x^{2},x^{3},v\right) dx^{3},  \label{4solrf1} \\
~^{\chi }\delta y &=&dy+n_{2}\left( x^{2},x^{3},v,\chi \right)
[dx^{2}+dx^{3}].  \notag
\end{eqnarray}%
Such solutions describe in general form the Ricci flows of nonholonomic
Einstein spaces constrained to relate in a mutually compatible form the
evolution of horizontal part of metric, $\varpi (x^{2},x^{3},\chi ),$ with
the evolution of N--connection coefficients $n_{2}=n_{3}=n_{2}\left(
x^{2},x^{3},v,\chi \right).$ We have to impose certain boundary/ initial
conditions for $\chi =0,$ beginning with an explicit solution of the
Einstein equations, in order to define the integration functions and state
an evolution scenario for such classes of metrics and connections.

\subsection{4D and 5D Einstein foliations and Ricci flows}

The method of constructing 5D solutions can be restricted to generate 4D
nonholonomic configurations and generic off--diagonal solutions in general
relativity. In order to consider reductions $5D\rightarrow 4D$ for the
ansatz (\ref{4ans5d}), we can eliminate from the formulas the variable $%
x^{1} $ and consider a 4D space $\mathbf{V}^{4}$ (parametrized by local
coordinates $\left( x^{2},x^{3},v,y^{5}\right) )$ trivially embedded into a
5D spacetime $\mathbf{V}$ (parametrized by local coordinates $\left(
x^{1},x^{2},x^{3},v,y^{5}\right) $ with $g_{11}=\pm 1,g_{1\widehat{\alpha }%
}=0,\widehat{\alpha }=2,3,4,5).$ In this case, there are possible \ 4D
conformal and anholonomic transforms depending only on variables $\left(
x^{2},x^{3},v\right) $ of a 4D metric $g_{\widehat{\alpha }\widehat{\beta }%
}\left( x^{2},x^{3},v\right) $ of arbitrary signature. To emphasize that
some coordinates are stated just for a 4D space we might use ''hats'' on the
Greek indices, $\widehat{\alpha },\widehat{\beta },...$ \ and on the Latin
indices from the middle of the alphabet, $\widehat{i},\widehat{j},...=2,3;$
local coordinates on $\mathbf{V}^{4}$ are parametrized $u^{\widehat{\alpha }%
}=\left( x^{\widehat{i}},y^{a}\right) =\left(
x^{2},x^{3},y^{4}=v,y^{5}\right) ,$ for $a,b,...=4,5.$ The ansatz
\begin{equation}
\mathbf{g}=g_{2}\ dx^{2}\otimes dx^{2}+g_{3}\ dx^{3}\otimes dx^{3}+h_{4}\
\delta v\otimes \delta v+h_{5}\ \delta y^{5}\otimes \delta y^{5},
\label{4dmetric4}
\end{equation}%
is written with respect to the anholonomic co--frame $\left( dx^{\widehat{i}%
},\delta v,\delta y^{5}\right) ,$ where
\begin{equation}
\delta v=dv+w_{\widehat{i}}dx^{\widehat{i}}\mbox{ and }\delta
y^{5}=dy^{5}+n_{\widehat{i}}dx^{\widehat{i}}  \label{4ddif4}
\end{equation}%
is the dual of $\left( \delta _{\widehat{i}},\partial _{4},\partial
_{5}\right) ,$ for
\begin{equation}
\delta _{\widehat{i}}=\partial _{\widehat{i}}+w_{\widehat{i}}\partial
_{4}+n_{\widehat{i}}\partial _{5},  \label{4dder4}
\end{equation}%
and the coefficients are necessary smoothly class functions of type:
\begin{eqnarray}
g_{\widehat{j}} &=&g_{\widehat{j}}(x^{\widehat{k}}),h_{4,5}=h_{4,5}(x^{%
\widehat{k}},v),  \notag \\
w_{\widehat{i}} &=&w_{\widehat{i}}(x^{\widehat{k}},v),n_{\widehat{i}}=n_{%
\widehat{i}}(x^{\widehat{k}},v);~\widehat{i},\widehat{k}=2,3.  \notag
\end{eqnarray}

In the 4D case, a source of type (\ref{4sdiag}) should be considered without
the component $\Upsilon _{1}^{1}$ in the form
\begin{equation}
\mathbf{\Upsilon }_{\widehat{\beta }}^{\widehat{\alpha }}=diag[\Upsilon
_{2}^{2}=\Upsilon _{3}^{3}=\Upsilon _{2}(x^{\widehat{k}},v),\ \Upsilon
_{4}^{4}=\Upsilon _{5}^{5}=\Upsilon _{4}(x^{\widehat{k}})].  \label{4sdiag4}
\end{equation}%
The Einstein equations with sources of type (\ref{4sdiag4}) for the
canonical d--con\-nec\-ti\-on (\ref{4candcon}) defined by the ansatz (\ref%
{4dmetric4}) transform into a system of nonlinear partial differential
equations very similar to (\ref{4ep1a})--(\ref{4ep4a}). The difference for
the 4D equations is that the coordinate $x^{1}$ is not contained into the
equations and that the indices of type $i,j,..=1,2,3$ must be changed into
the corresponding indices $\widehat{i},\widehat{j},..=2,3.$ The generated
classes of 4D solutions are defined almost by the same formulas (\ref%
{4gensol1}), (\ref{4gensol1w}) and (\ref{4gensol1n}).

Now we describe how the coefficients of an ansatz (\ref{4dmetric4}) defining
an exact vacuum solution for a canonical d--connection can be constrained to
generate a vacuum solution in Einstein gravity: We start with the
conditions\ (\ref{4cond3}) written (for our ansatz) in the form%
\begin{eqnarray}
\frac{\partial h_{4}}{\partial x^{\widehat{k}}}-w_{\widehat{k}}h_{4}^{\ast
}-2w_{\widehat{k}}^{\ast }h_{4} &=&0,  \label{4cond3a} \\
\frac{\partial h_{5}}{\partial x^{\widehat{k}}}-w_{\widehat{k}}h_{5}^{\ast }
&=&0,  \label{4cond3b} \\
n_{\widehat{k}}^{\ast }h_{5} &=&0.  \label{4cond3c}
\end{eqnarray}%
These equations for nontrivial values of $w_{\widehat{k}}$ and $n_{\widehat{k%
}}$ constructed for some defined values of \ $h_{4}$ and $h_{5}$ must be
compatible with the equations (\ref{4ep2a})--(\ref{4ep4a}) for $\Upsilon
_{2}=0.$ One can be taken nonzero values for $w_{\widehat{k}}$ in (\ref%
{4ep3a}) if and only if $\alpha _{\widehat{i}}=0$ because the the equation (%
\ref{4ep2a}) imposes the condition $\beta =0.$ This is possible, for the
sourceless case and $h_{5}^{\ast }\neq 0,$ if and only if
\begin{equation}
\phi =\ln \left| h_{5}^{\ast }/\sqrt{|h_{4}h_{5}|}\right| =const,
\label{4eq23}
\end{equation}%
see formula (\ref{4coefa}). A very general class of solutions of equations (%
\ref{4cond3a}), (\ref{4cond3b}) and (\ref{4eq23}) can be represented in the
form
\begin{eqnarray}
h_{4} &=&\epsilon _{4}h_{0}^{2}\ \left( b^{\ast }\right) ^{2},h_{5}=\epsilon
_{5}(b+b_{0})^{2},  \label{4lcc2} \\
w_{\widehat{k}} &=&\left( b^{\ast }\right) ^{-1}\frac{\partial (b+b_{0})}{%
\partial x^{\widehat{k}}},  \notag
\end{eqnarray}%
where $h_{0}=const$ and $b=b(x^{\widehat{k}},v)$ is any function for which $%
b^{\ast }\neq 0$ and $b_{0}=b_{0}(x^{\widehat{k}})$ is an arbitrary
integration function.

The next step is to satisfy the integrability conditions (\ref{4fols})
defining a foliated spacetimes provided with metric and N--connection and
d--connection structures \cite{4vesnc,4vsgg,4esv} (we note that (pseudo)
Riemannian foliations are considered in a different manner in Ref. \cite%
{4bejf}) for the so--called Schouten -- Van Kampen and Vranceanu connections
not subjected to the condition to generate Einstein spaces). It is very easy
to show that there are nontrivial solutions of the constraints (\ref{4fols})
which for the ansatz (\ref{4dmetric4}) are written in the form
\begin{eqnarray}
w_{2}^{\prime }-w_{3}^{\bullet }+w_{3}w_{2}^{\ast }-w_{2}w_{3}^{\ast } &=&0,
\label{4aux31} \\
n_{2}^{\prime }-n_{3}^{\bullet }+w_{3}n_{2}^{\ast }-w_{2}n_{3}^{\ast } &=&0.
\notag
\end{eqnarray}%
We solve these equations for $n_{2}^{\ast }=n_{3}^{\ast }=0$ if we take any
two functions $n_{2,3}(x^{\widehat{k}})$ satisfying
\begin{equation}
n_{2}^{\prime }-n_{3}^{\bullet }=0  \label{4aux31b}
\end{equation}%
(this is possible by a particular class of integration functions in (\ref%
{4gensol1n}) when $n_{\widehat{k}[2]}\left( x^{\widehat{i}}\right) =0$ and $%
n_{\widehat{k}[1]}\left( x^{\widehat{i}}\right) $ are constraint to satisfy
just the conditions (\ref{4aux31b})). Then we can consider any function $%
b(x^{\widehat{k}},v)$ for which $w_{\widehat{k}}=\left( b^{\ast }\right)
^{-1}$ $\partial _{\widehat{k}}(b+b_{0})$ solve the equation (\ref{4aux31}).
In a more particular case, one can be constructed solutions for any $%
b(x^{3},v),b^{\ast }\neq 0,$ and $n_{2}=0$ and $n_{3}=n_{3}(x^{3},v)$ (or,
inversely, for any $n_{2}=n_{2}(x^{2},v)$ and $n_{3}=0).$ We also note  that
the conditions (\ref{4coef0}) are solved in a straightforward form by the
ansatz (\ref{4dmetric4}).

We conclude that for any sets of coefficients
\begin{equation*}
h_{4}(x^{\widehat{k}},v),\ h_{5}(x^{\widehat{k}},v),\ w_{\widehat{k}}(x^{%
\widehat{k}},v),\ n_{2,3}(x^{\widehat{k}})
\end{equation*}
respectively generated by functions $b(x^{\widehat{k}},v)$ and $n_{\widehat{k%
}[1]}\left( x^{\widehat{i}}\right) ,$ see (\ref{4lcc2}), and satisfying (\ref%
{4aux31b}), the generic off--diagonal metric (\ref{4dmetric4}) possess the
same coefficients both for the Levi Civita and canonical d--connection being
satisfied the conditions (\ref{4cond1}) of equality of the Einstein tensors.
Here we note that any 2D metric can be written in a conformally flat form,
i. e. we can chose such local coordinates when
\begin{equation*}
g_{2}(dx^{2})^{2}+g_{3}(dx^{3})^{2}=e^{\psi (x^{\widehat{i}})}\left[
\epsilon _{\widehat{2}}(dx^{\widehat{2}})^{2}+\epsilon _{\widehat{3}}(dx^{%
\widehat{3}})^{2}\right] ,
\end{equation*}%
for signatures $\epsilon _{\widehat{k}}=\pm 1,$ in (\ref{4dmetric4}).

Summarizing the results of this section, we can write down the generic
off--diagonal metric (it is a 4D dimensional reduction of (\ref{4gensol1}))
\begin{eqnarray}
\ _{\shortmid }^{\circ }\mathbf{g} &=&e^{\psi (x^{2},x^{3})}\left[ \epsilon
_{2}\ dx^{2}\otimes dx^{2}+\epsilon _{3}\ dx^{3}\otimes dx^{3}\right]
\label{4ds} \\
&&+\epsilon _{4}h_{0}^{2}\ \left[ b^{\ast }\left( x^{i},v\right) \right]
^{2}\ \delta v\otimes \delta v  \notag \\
&&+\epsilon _{5}\left[ b\left( x^{2},x^{3},v\right) -b_{0}(x^{2},x^{3})%
\right] ^{2}\ \delta y^{5}\otimes \delta y^{5},  \notag \\
\delta v &=&dv+w_{2}\left( x^{2},x^{3},v\right) dx^{2}+w_{3}\left(
x^{2},x^{3},v\right) dx^{3},  \notag \\
\ \delta y^{5} &=&dy^{5}+n_{2}\left( x^{2},x^{3}\right) dx^{2}+n_{3}\left(
x^{2},x^{3}\right) dx^{3},  \notag
\end{eqnarray}%
defining vacuum exact solutions in general relativity if the coefficients
are restricted to solve the equations
\begin{eqnarray}
\epsilon _{2}\psi ^{\bullet \bullet }+\epsilon _{3}\psi ^{^{\prime \prime }}
&=&0,  \label{4cond5} \\
w_{2}^{\prime }-w_{3}^{\bullet }+w_{3}w_{2}^{\ast }-w_{2}w_{3}^{\ast } &=&0,
\notag \\
n_{2}^{\prime }-n_{3}^{\bullet } &=&0,  \notag
\end{eqnarray}%
for $w_{2}=\left( b^{\ast }\right) ^{-1}(b+b_{0})^{\bullet }$ and $%
w_{3}=\left( b^{\ast }\right) ^{-1}(b+b_{0})^{\prime },$ where, for
instance, $n_{3}^{\bullet }=\partial _{2}n_{3}$ and $n_{2}^{\prime
}=\partial _{3}n_{2}.$

We can generalize (\ref{4ds}) similarly to (\ref{4gensol1}) in order to
generate solutions for nontrivial sources (\ref{4sdiag4}). In general, they
will contain nontrivial anholonomically induced torsions. Such
configurations may be restricted to the case of Levi Civita connection by
solving the constraints (\ref{4cond3a})--(\ref{4cond3c}) in order to be
compatible with the equations (\ref{4ep2a}) and (\ref{4ep3a}) for the
coefficients $\alpha _{\widehat{i}}$ and $\beta $ computed for $h_{5}^{\ast
}\neq 0$ and $\ln \left| h_{5}^{\ast }/\sqrt{|h_{4}h_{5}|}\right| =\phi
(x^{2},x^{3},v)\neq const,$ see formula (\ref{4coefa}), resulting in more
general conditions than (\ref{4eq23}) and (\ref{4lcc2}). Roughly speaking,
all such coefficients are generated by any $h_{4}$ (or $h_{5}$) defined from
(\ref{4ep3a}) for prescribed values $h_{5}$ (or $h_{5}$) and $\Upsilon
_{2}(x^{\widehat{k}},v).$ The existence of a nontrivial matter source of
type (\ref{4sdiag4}) does not change the condition $n_{\widehat{k}}^{\ast
}=0,$ see (\ref{4cond3c}), necessary for extracting torsionless
configurations. This mean that we have to consider only trivial solutions of
(\ref{4ep4a}) when two functions $n_{\widehat{k}}=n_{\widehat{k}%
}(x^{2},x^{3})$ are subjected to the condition (\ref{4aux31}). We conclude
that this class of exact solutions of the Einstein equations with nontrivial
sources (\ref{4sdiag4}), in general relativity, is defined by the ansatz
\begin{eqnarray}
\ _{\shortmid }^{\circ }\mathbf{g} &=&e^{\psi (x^{2},x^{3})}\left[ \epsilon
_{2}\ dx^{2}\otimes dx^{2}+\epsilon _{3}\ dx^{3}\otimes dx^{3}\right] +
\label{4es4s} \\
&&h_{4}\left( x^{2},x^{3},v\right) \ \delta v\otimes \delta v+h_{5}\left(
x^{2},x^{3},v\right) \ \delta y^{5}\otimes \delta y^{5},  \notag \\
\delta v &=&dv+w_{2}\left( x^{2},x^{3},v\right) dx^{2}+w_{3}\left(
x^{2},x^{3},v\right) dx^{3},  \notag \\
\ \delta y^{5} &=&dy^{5}+n_{2}\left( x^{2},x^{3}\right) dx^{2}+n_{3}\left(
x^{2},x^{3}\right) dx^{3},  \notag
\end{eqnarray}%
where the coefficients satisfy the conditions
\begin{eqnarray}
\epsilon _{2}\psi ^{\bullet \bullet }+\epsilon _{3}\psi ^{^{\prime \prime }}
&=&\Upsilon _{2}  \notag \\
h_{5}^{\ast }\phi /h_{4}h_{5} &=&\Upsilon _{2},  \label{4ep2b} \\
w_{2}^{\prime }-w_{3}^{\bullet }+w_{3}w_{2}^{\ast }-w_{2}w_{3}^{\ast } &=&0,
\notag \\
n_{2}^{\prime }-n_{3}^{\bullet } &=&0,  \notag
\end{eqnarray}%
for $w_{\widehat{i}}=\partial _{\widehat{i}}\phi /\phi ^{\ast },$ see (\ref%
{4coefa}), being compatible with (\ref{4cond3a}) and (\ref{4cond3b}), for
given sources $\Upsilon _{4}(x^{\widehat{k}})$ and $\Upsilon _{2}(x^{%
\widehat{k}},v).$ We emphasize that the second equation in (\ref{4ep2b})
relates two functions $h_{4}$ and $h_{5}.$ In references \cite%
{4vhep,4vt,4vp,4vs,4vesnc}, we investigated a number of configurations with
nontrivial two and three dimensional solitons, reductions to the Riccati or
Abbel equation, defining off--diagonal deformations of the black hole,
wormhole or Taub NUT spacetimes. Those solutions where constructed to be
with trivial or nontrivial torsions but if the coefficients of the ansatz (%
\ref{4es4s}) are restricted to satisfy the conditions (\ref{4ep2b}) in a
compatible form with (\ref{4cond3a}) and (\ref{4cond3b}), for sure, such
metrics will solve the Einstein equations for the Levi Civita connection. We
emphasize that the ansatz (\ref{4es4s}) defines Einstein spaces with a
cosmological constant $\lambda $ if we put $\Upsilon _{2}=\Upsilon
_{4}=\lambda $ in (\ref{4ep2b}).

Let us formulate the conditions when families of metrics (\ref{4es4s})
subjected to the conditions (\ref{4ep2b}) will define exact solutions of the
Ricci flows of usual Einstein spaces (for the Levi Civita connection). We
consider the ansatz
\begin{eqnarray}
\ _{\shortmid }^{\lambda }\mathbf{g}(\chi ) &=&e^{\psi (x^{2},x^{3},\chi )}%
\left[ \epsilon _{2}\ dx^{2}\otimes dx^{2}+\epsilon _{3}\ dx^{3}\otimes
dx^{3}\right] +  \label{4slerf1e} \\
&&h_{4}\left( x^{2},x^{3},v,\chi \right) \ \delta v\otimes \delta
v+h_{5}\left( x^{2},x^{3},v,\chi \right) \ ^{\chi }\delta y^{5}\otimes \
^{\chi }\delta y^{5},  \notag \\
\delta v &=&dv+w_{2}\left( x^{2},x^{3},v\right) dx^{2}+w_{3}\left(
x^{2},x^{3},v\right) dx^{3},  \notag \\
\ \ ^{\chi }\delta y^{5} &=&dy^{5}+n_{2}\left( x^{2},x^{3},\chi \right)
[dx^{2}+dx^{3}],  \notag
\end{eqnarray}%
which is a subfamily of (\ref{4solrf1}), when $\varpi =e^{\psi
(x^{2},x^{3},\chi )}$ and $n_{2}=n_{3}$ does not depend on variable $v$ and
the coefficients satisfy the conditions (\ref{4rfea1}) and (\ref{4rfea2}), \
when $n_{[2]}=0$ but $n_{[1]}$ \ can be nontrivial in \ (\ref{4rfea3}), and
(additionally)
\begin{eqnarray}
\epsilon _{2}\psi ^{\bullet \bullet }(\chi )+\epsilon _{3}\psi ^{^{\prime
\prime }}(\chi ) &=&\lambda ,  \notag \\
h_{5}^{\ast }\phi /h_{4}h_{5} &=&\lambda ,  \label{4ep2be} \\
w_{2}^{\prime }-w_{3}^{\bullet }+w_{3}w_{2}^{\ast }-w_{2}w_{3}^{\ast } &=&0,
\notag \\
n_{2}^{\prime }(\chi )-n_{2}^{\bullet }(\chi ) &=&0,  \notag
\end{eqnarray}%
for $w_{\widehat{i}}=\partial _{\widehat{i}}\phi /\phi ^{\ast },$ see (\ref%
{4coefa}), being compatible with (\ref{4cond3a}) and (\ref{4cond3b}), for
given sources $\Upsilon _{4}=\lambda $ and $\Upsilon _{2}=\lambda .$ The
family of metrics (\ref{4slerf1e}) define a self--consistent evolution as a
class of general solutions of the Ricci flow equations (\ref{4eq1})--(\ref%
{4eq3}) transformed equivalently in the form (\ref{4ep1a})--(\ref{4ep4a}).
The additional constraints (\ref{4ep2be}) define an integral subvariety
(foliation) of (\ref{4solrf1}) when the evolution is selected for the Levi
Civita connection.

\section{Nonholonomic and Parametric Transforms}

\label{ssndm} Anholonomic deformations can be defined for any primary metric
and frame structures on a spacetime $\mathbf{V}$ (as a matter of principle,
the primary metric can be not a solution of the gravitational field
equations). Such deformations always result in a target spacetime possessing
one Killing vector symmetry if the last one is constrained to satisfy the
vacuum Einstein equations for the canonical d--connection, or for the Levi
Civita connection. For such target spacetimes, we can always apply a
parametric transform and generate a set of generic off--diagonal solutions
labelled by a parameter $\theta $ (\ref{4ger1t}). There are possible
constructions when the anholonomic frame transforms are applied to a family
of metrics generated by the parametric method as new exact solutions of the
vacuum Einstein equations, but such primary metrics have to be parametrized
by certain type ansatz admitting anholonomic transforms to other classes of
exact solutions. Additional constraints and parametrizations are necessary
for generating exact solutions of holonomic or nonholonomic Ricci flow
equations.

\subsection{Deformations and frame parametrizations}

Let us consider a $(n+m)$--dimensional manifold (spacetime) $\mathbf{V},$ $%
n\geq 2,m\geq 1,$ enabled with a metric structure $\mathbf{\check{g}}=\check{%
g}\oplus _{N}\ \check{h}$ distinguished in the form
\begin{eqnarray}
\mathbf{\check{g}} &=&\check{g}_{i}(u)(dx^{i})^{2}+\check{h}_{a}(u)(\mathbf{%
\check{c}}^{a})^{2},  \label{4m1} \\
\mathbf{\check{c}}^{a} &=&dy^{a}+\check{N}_{i}^{a}(u)dx^{i}.  \notag
\end{eqnarray}%
The local coordinates are parametrized $u=(x,y)=\{u^{\alpha
}=(x^{i},y^{a})\},$ for the indices of type $i,j,k,...=1,2,...,n$ (in brief,
horizontal, or h--indices/ components) and $a,b,c,...=n+1,n+2,...n+m$
(vertical, or v--indices/ components). We suppose that, in general, the
metric (\ref{4m1}) is not a solution of the Einstein equations but can be
nonholonomically deformed in order to generate exact solutions. The
coefficients $\check{N}_{i}^{a}(u)$ from (\ref{4m1}) state a conventional $%
\left( n+m\right) $--splitting $\oplus _{\check{N}}$ in any point $u\in
\mathbf{V}$ and define a class of 'N--adapted' local bases
\begin{equation}
\mathbf{\check{e}}_{\alpha }=\left( \mathbf{\check{e}}_{i}=\frac{\partial }{%
\partial x^{i}}-\check{N}_{i}^{a}(u)\ \frac{\partial }{\partial y^{a}},e_{a}=%
\frac{\partial }{\partial y^{a}}\right)  \label{4b1}
\end{equation}%
and local dual bases (co--frames) $\mathbf{\check{c}}=(c,\check{c}),$ when
\begin{equation}
\mathbf{\check{c}}^{\alpha }=\left( c^{j}=dx^{i},\mathbf{\check{c}}%
^{b}=dy^{b}+\check{N}_{i}^{b}(u)\ dx^{i}\right) ,  \label{4cb1}
\end{equation}%
for $\mathbf{\check{c}\rfloor \ \check{e}=I,}$ i.e. $\mathbf{\check{e}}%
_{\alpha }\mathbf{\rfloor }$ $\mathbf{\check{c}}^{\beta }=\delta _{\alpha
}^{\beta },$ where the inner product is denoted by '$\mathbf{\rfloor }$' and
the Kronecker symbol is written $\delta _{\alpha }^{\beta }.$ The frames (%
\ref{4b1}) satisfy the nonholonomy (equivalently, anholonomy) relations
\begin{equation*}
\mathbf{\check{e}}_{\alpha }\mathbf{\check{e}}_{\beta }-\mathbf{\check{e}}%
_{\beta }\mathbf{\check{e}}_{\alpha }=\mathbf{\check{w}}_{\alpha \beta
}^{\gamma }\mathbf{\check{e}}_{\gamma }
\end{equation*}%
with nontrivial anholonomy coefficients
\begin{eqnarray}
\mathbf{\check{w}}_{ji}^{a} &=&-\mathbf{\check{w}}_{ij}^{a}=\mathbf{\check{%
\Omega}}_{ij}^{a}\doteqdot \mathbf{\check{e}}_{j}\left( \check{N}%
_{i}^{a}\right) -\mathbf{\check{e}}_{i}\left( \check{N}_{j}^{a}\right) ,
\label{4anhc} \\
\mathbf{\check{w}}_{ia}^{b} &=&-\mathbf{\check{w}}_{ai}^{b}=e_{a}(\check{N}%
_{j}^{b}).  \notag
\end{eqnarray}

A metric $\mathbf{g}=g\oplus _{N}h$ parametrized in the form
\begin{eqnarray}
\mathbf{g} &=&g_{i}(u)(c^{i})^{2}+g_{a}(u)(\mathbf{c}^{a}),  \label{4m2} \\
\mathbf{c}^{a} &=&dy^{a}+N_{i}^{a}(u)dx^{i}  \notag
\end{eqnarray}%
is a nonhlonomic transform (deformation), preserving the $(n+m)$--splitting,
of the metric, $\mathbf{\check{g}}=\check{g}\oplus _{\check{N}}\ \check{h}$
if the coefficients of (\ref{4m1}) and (\ref{4m2}) are related by formulas
\begin{equation}
g_{i}=\eta _{i}(u)\ \check{g}_{i},\ h_{a}=\eta _{a}(u)\ \check{h}_{a}%
\mbox{
and }N_{i}^{a}=\eta _{i}^{a}(u)\check{N}_{i}^{a},  \label{4polf}
\end{equation}%
where the summation rule is not considered for the indices of gravitational
'polarizations' $\eta _{\alpha }=(\eta _{i},\eta _{a})$ \ and $\eta _{i}^{a}$
in (\ref{4polf}). For nontrivial values of $\eta _{i}^{a}(u),$ the
nonholonomic frames (\ref{4b1}) and (\ref{4cb1}) transform correspondingly
into
\begin{equation}
\mathbf{e}_{\alpha }=\left( \mathbf{e}_{i}=\frac{\partial }{\partial x^{i}}%
-N_{i}^{a}(u)\ \frac{\partial }{\partial y^{a}},e_{a}=\frac{\partial }{%
\partial y^{a}}\right)  \label{4b1a}
\end{equation}%
and
\begin{equation}
\mathbf{c}^{\alpha }=\left( c^{j}=dx^{i},\mathbf{c}^{a}=dy^{a}+N_{i}^{a}(u)\
dx^{i}\right)  \label{4cb1a}
\end{equation}%
with the anholonomy coefficients $\mathbf{W}_{\alpha \beta }^{\gamma }$
defined by $N_{i}^{a}$ (\ref{4anhncc}).

We emphasize that in order to generate exact solutions, the gravitational
'polarizations' $\eta _{\alpha }=(\eta _{i},\eta _{a})$ \ and $\eta _{i}^{a}$
in (\ref{4polf}) are not arbitrary functions but restricted in a such form
that the values
\begin{eqnarray}
\pm 1 &=&\eta _{1}(u^{\alpha })\ \check{g}_{1}(u^{\alpha }),\
\label{4polf1} \\
g_{2}(x^{2},x^{3}) &=&\eta _{2}(u^{\alpha })\ \check{g}_{2}(u^{\alpha }),\
g_{3}(x^{2},x^{3})=\eta _{3}(u^{\alpha })\ \check{g}_{3}(u^{\alpha }),
\notag \\
h_{4}(x^{i},v) &=&\eta _{4}(u^{\alpha })\ \check{h}_{4}(u^{\alpha }),\
h_{5}(x^{i},v)=\eta _{5}(u^{\alpha })\ \check{h}_{5}(u^{\alpha }),  \notag \\
w_{i}(x^{i},v) &=&\eta _{i}^{4}(u^{\alpha })\check{N}_{i}^{4}(u^{\alpha }),\
n_{i}(x^{i},v)=\eta _{i}^{5}(u^{\alpha })\check{N}_{i}^{5}(u^{\alpha }),
\notag
\end{eqnarray}%
define an ansatz of type (\ref{4gensol1}), or (\ref{4ds}) (for vacuum
configurations) and (\ref{4es4s}) for nontrivial matter sources $\Upsilon
_{2}(x^{2},x^{3},v)$ and $\Upsilon _{4}(x^{2},x^{3}).$

Any nonholonomic deformation%
\begin{equation}
\mathbf{\check{g}}=\check{g}\oplus _{\check{N}}\ \check{h}\longrightarrow \
\mathbf{g}=g\oplus _{N}h  \label{4nfd}
\end{equation}%
can be described by two frame matrices of type (\ref{4vt1}),
\begin{equation}
\mathbf{\check{A}}_{\alpha }^{\ \underline{\alpha }}(u)=\left[
\begin{array}{cc}
\delta _{i}^{\ \underline{i}} & -\check{N}_{j}^{b}\delta _{b}^{\ \underline{a%
}} \\
0 & \delta _{a}^{\ \underline{a}}%
\end{array}%
\right] ,  \label{4vtn1}
\end{equation}%
generating the d--metric $\mathbf{\check{g}}_{\alpha \beta }=\mathbf{\check{A%
}}_{\alpha }^{\ \underline{\alpha }}\mathbf{\check{A}}_{\beta }^{\
\underline{\beta }}\check{g}_{\underline{\alpha }\underline{\beta }},$ see
formula (\ref{4fmt}), and
\begin{equation}
\mathbf{A}_{\alpha }^{\ \underline{\alpha }}(u)=\left[
\begin{array}{cc}
\sqrt{|\eta _{i}|}\delta _{i}^{\ \underline{i}} & -\eta _{i}^{a}\check{N}%
_{j}^{b}\delta _{b}^{\ \underline{a}} \\
0 & \sqrt{|\eta _{a}|}\delta _{a}^{\ \underline{a}}%
\end{array}%
\right] ,  \label{4vtn2}
\end{equation}%
generating the d--metric $\mathbf{g}_{\alpha \beta }=\mathbf{A}_{\alpha }^{\
\underline{\alpha }}\mathbf{A}_{\beta }^{\ \underline{\beta }}\check{g}_{%
\underline{\alpha }\underline{\beta }}$ (\ref{4polf1}).

If the metric and N--connection coefficients (\ref{4polf}) are stated to be
those from an ansatz (\ref{4gensol1}) (or (\ref{4ds})), we should write $\
^{\circ }\mathbf{g}=g\oplus _{N}h$ (or $\ _{\shortmid }^{\circ }\mathbf{g}%
=g\oplus _{N}h$) and say that the metric $\mathbf{\check{g}}=\check{g}\oplus
_{N}\ \check{h}$ (\ref{4m1}) was nonholonomically deformed in order to
generate an exact solution of the Einstein equations for the canonical
d--connection (or, in a restricted case, for the Levi Civita connection). In
general, such metrics have very different geometrical and physical
properties. Nevertheless, at least for some classes of 'small' nonsingular
nonholonomic deformations, it is possible to preserve a similar physical
interpretation by introducing small polarizations of metric coefficients and
deformations of existing horizons, not changing the singular structure of
curvature tensors. Explicit examples are constructed in Ref. \cite{4nhrf05}.

\subsection{The Geroch transforms as parametric nonholono\-mic
de\-forma\-ti\-ons}

We note that any metric $\ _{\shortmid }^{\circ }g_{\alpha \beta }$ defining
an exact solution of the vacuum Einstein equations can be represented in the
form (\ref{4m1}). Then, any metric $\ _{\shortmid }^{\circ }\widetilde{g}%
_{\alpha \beta }(\theta )$ (\ref{4ger1t}) from a family of new solutions
generated by the first type parametric transform can be written as (\ref{4m2}%
) and related via certain polarization functions of type (\ref{4polf}), in
the parametric case depending on parameter $\theta ,$ i.e. $\ \eta _{\alpha
}(\theta )=(\eta _{i}(\theta ),\eta _{a}(\theta ))$ \ and $\eta
_{i}^{a}(\theta ).$ Roughly speaking, any parametric transform can be
represented as a generalized class of anholonomic frame transforms
additionally parametrized by $\theta $ and adapted to preserve the $(n+m)$%
--splitting structure.\footnote{%
It should be emphasized that such constructions are not trivial, for usual
coordinate transforms, if at least one of the primary or target metrics is
generic off--diagonal.} The corresponding frame transforms (\ref{4qe}) and (%
\ref{4qef}) are parametrized, respectively, by matrices of type (\ref{4vtn1}%
) and (\ref{4vtn2}), also ''labelled'' by $\theta .$ Such nonholonomic
parametric deformations
\begin{equation}
\ _{\shortmid }^{\circ }\mathbf{g}=\ _{\shortmid }^{\circ }g\oplus _{%
\check{N}}\ _{\shortmid }^{\circ }h\longrightarrow \ \ _{\shortmid }^{\circ }%
\widetilde{\mathbf{g}}(\theta )=\ _{\shortmid }^{\circ }\widetilde{g}(\theta
)\oplus _{N(\theta )}\ _{\shortmid }^{\circ }\widetilde{h}(\theta )
\label{4gfd}
\end{equation}%
are defined by the frame matrices,
\begin{equation}
\ _{\shortmid }^{\circ }\mathbf{A}_{\alpha }^{\ \underline{\alpha }}(u)=%
\left[
\begin{array}{cc}
\delta _{i}^{\ \underline{i}} & -\ _{\shortmid }^{\circ }N_{j}^{b}(u)\delta
_{b}^{\ \underline{a}} \\
0 & \delta _{a}^{\ \underline{a}}%
\end{array}%
\right] ,  \label{4qep}
\end{equation}%
generating the d--metric $\ _{\shortmid }^{\circ }\mathbf{g}_{\alpha \beta
}=\ _{\shortmid }^{\circ }\mathbf{A}_{\alpha }^{\ \underline{\alpha }}\
_{\shortmid }^{\circ }\mathbf{A}_{\beta }^{\ \underline{\beta }}\
_{\shortmid }^{\circ }g_{\underline{\alpha }\underline{\beta }}$ and
\begin{equation}
\widetilde{\mathbf{A}}_{\alpha }^{\ \underline{\alpha }}(u,\theta )=\left[
\begin{array}{cc}
\sqrt{|\eta _{i}(u,\theta )|}\delta _{i}^{\ \underline{i}} & -\eta
_{i}^{a}(u,\theta )\ _{\shortmid }^{\circ }N_{j}^{b}(u)\delta _{b}^{\
\underline{a}} \\
0 & \sqrt{|\eta _{a}(u,\theta )|}\delta _{a}^{\ \underline{a}}%
\end{array}%
\right] ,  \label{4qefp}
\end{equation}%
generating the d--metric $\ _{\shortmid }^{\circ }\widetilde{\mathbf{g}}%
_{\alpha \beta }(\theta )=\widetilde{\mathbf{A}}_{\alpha }^{\ \underline{%
\alpha }}\widetilde{\mathbf{A}}_{\beta }^{\ \underline{\beta }}\ _{\shortmid
}^{\circ }g_{\underline{\alpha }\underline{\beta }}.$ Using the matrices (%
\ref{4qep}) and (\ref{4qefp}), we can compute the matrix of parametric
transforms%
\begin{equation}
\widetilde{\mathbf{B}}_{\alpha }^{\ \alpha ^{\prime }}=\widetilde{\mathbf{A}}%
_{\alpha }^{\ \underline{\alpha }}\ \ _{\shortmid }^{\circ }\mathbf{A}_{%
\underline{\alpha }}^{\ \alpha ^{\prime }},  \label{4mgt1}
\end{equation}%
like in (\ref{4mgt}), but for ''boldfaced' objects, where $\ _{\shortmid
}^{\circ }\mathbf{A}_{\underline{\alpha }}^{\ \alpha ^{\prime }}$ is inverse
to $\ _{\shortmid }^{\circ }\mathbf{A}_{\alpha ^{\prime }}^{\ \underline{%
\alpha }},$ \footnote{%
we use a ''boldface'' symbol because in this case the constructions are
adapted to a $(n+m)$--splitting} and define the target set of metrics in the
form
\begin{equation*}
\ _{\shortmid }^{\circ }\widetilde{\mathbf{g}}_{\alpha \beta }=\widetilde{%
\mathbf{B}}_{\alpha }^{\ \alpha ^{\prime }}(u,\theta )\ \widetilde{\mathbf{B}%
}_{\beta }^{\ \beta ^{\prime }}(u,\theta )\ _{\shortmid }^{\circ }\mathbf{g}%
_{\alpha ^{\prime }\beta ^{\prime }}.
\end{equation*}

There are two substantial differences from the case of usual anholonomic
frame transforms (\ref{4nfd}) and the case of parametric deformations (\ref%
{4gfd}). The first one is that the metric $\mathbf{\check{g}}$ was not
constrained to be an exact solution of the Einstein equations like it was
required for $\ _{\shortmid }^{\circ }\mathbf{g.}$ The second one is that
even $\ \mathbf{g}$ can be restricted to be an exact vacuum solution,
generated by a special type of deformations (\ref{4polf1}), in order to get
an ansatz of type (\ref{4ds}), an arbitrary metric from a family of
solutions $\ _{\shortmid }^{\circ }\widetilde{\mathbf{g}}_{\alpha \beta
}(\theta )$ will not be parametrized in a form that the coefficients will
satisfy the conditions (\ref{4cond5}). Nevertheless, even in such cases, we
can consider additional nonholonomic frame transforms when $\mathbf{\check{g}%
}$ is transformed into an exact solution and any particular metric from the
set $\left\{ \ _{\shortmid }^{\circ }\widetilde{\mathbf{g}}_{\alpha \beta
}(\theta )\right\} $ will be deformed into an exact solution defined by an
ansatz (\ref{4ds}) with additional dependence on $\theta .$

By superpositions of nonholonomic deformations, we can paramet\-ri\-ze a
solution formally constructed following by the parametric method (from a
primary solution depending on variables $x^{2},x^{3})$ in the form
\begin{eqnarray}
\ _{\shortmid }^{\circ }\mathbf{\tilde{g}(}\theta \mathbf{)} &=&e^{\psi
(x^{2},x^{3},\theta )}\left[ \epsilon _{2}\ dx^{2}\otimes dx^{2}+\epsilon
_{3}\ dx^{3}\otimes dx^{3}\right]  \label{4dst} \\
&&+\epsilon _{4}h_{0}^{2}\ \left[ b^{\ast }\left( x^{i},v,\theta \right) %
\right] ^{2}\ \delta v\otimes \delta v  \notag \\
&&+\epsilon _{5}\left[ b\left( x^{2},x^{3},v,\theta \right)
-b_{0}(x^{2},x^{3},\theta )\right] ^{2}\ \delta y^{5}\otimes \delta y^{5},
\notag \\
\delta v &=&dv+w_{2}\left( x^{2},x^{3},v,\theta \right) dx^{2}+w_{3}\left(
x^{2},x^{3},v,\theta \right) dx^{3},  \notag \\
\ \delta y^{5} &=&dy^{5}+n_{2}\left( x^{2},x^{3},\theta \right)
dx^{2}+n_{3}\left( x^{2},x^{3},\theta \right) dx^{3},  \notag
\end{eqnarray}%
with the coefficients restricted to solve the equations (\ref{4cond5}) but
depending additionally on parameter $\theta ,$
\begin{eqnarray}
\epsilon _{2}\psi ^{\bullet \bullet }\mathbf{(}\theta \mathbf{)}+\epsilon
_{3}\psi ^{^{\prime \prime }}\mathbf{(}\theta \mathbf{)} &=&0,
\label{4const6} \\
w_{2}^{\prime }\mathbf{(}\theta \mathbf{)}-w_{3}^{\bullet }\mathbf{(}\theta
\mathbf{)}+w_{3}w_{2}^{\ast }\mathbf{(}\theta \mathbf{)}-w_{2}\mathbf{(}%
\theta \mathbf{)}w_{3}^{\ast }\mathbf{(}\theta \mathbf{)} &=&0,  \notag \\
n_{2}^{\prime }\mathbf{(}\theta \mathbf{)}-n_{3}^{\bullet }\mathbf{(}\theta
\mathbf{)} &=&0,  \notag
\end{eqnarray}%
for $w_{2}\mathbf{(}\theta \mathbf{)}=\left( b^{\ast }\mathbf{(}\theta
\mathbf{)}\right) ^{-1}(b\mathbf{(}\theta \mathbf{)}+b_{0}\mathbf{(}\theta
\mathbf{)})^{\bullet }$ and $w_{3}=\left( b^{\ast }\mathbf{(}\theta \mathbf{)%
}\right) ^{-1}(b\mathbf{(}\theta \mathbf{)}+b_{0}\mathbf{(}\theta \mathbf{)}%
)^{\prime },$ where, for instance, $n_{3}^{\bullet }\mathbf{(}\theta \mathbf{%
)}=\partial _{2}n_{3}\mathbf{(}\theta \mathbf{)}$ and $n_{2}^{\prime
}=\partial _{3}n_{2}\mathbf{(}\theta \mathbf{)}.$

One should be noted that even, in general, any primary solution $\
_{\shortmid }^{\circ }\mathbf{g}$ can not be parametrized as an ansatz (\ref%
{4ds}), it is possible to define nonholonomic deformations to a such type
generic off--diagonal ansatz $\ _{\shortmid }^{\circ }\mathbf{\check{g}}$ $\
$or any $\mathbf{\check{g}},$ defined by an ansatz (\ref{4m1}), which in its
turn can be transformed into a metric of type (\ref{4dst}) without
dependence on $\theta .$\footnote{%
in our formulas we shall not point dependencies on coordinate variables if
that will not result in ambiguities}

Finally, we emphasize that in spite of the fact that both the parametric and
anholonomic frame transforms can be parametrized in very similar forms by
using frame transforms there is a criteria distinguishing one from another:
For a ''pure'' parametric transform, the matrix $\widetilde{\mathbf{B}}%
_{\alpha }^{\ \alpha ^{\prime }}(u,\theta )$ and related $\widetilde{\mathbf{%
A}}_{\alpha }^{\ \underline{\alpha }}\ $and$\ _{\shortmid }^{\circ }\mathbf{A%
}_{\underline{\alpha }}^{\ \alpha ^{\prime }}$ are generated by a solution
of the Geroch equations (\ref{4eq01}). If the ''pure'' nonholonomic
deformations, or their superposition with a parametric transform, are
introduced into consideration, the matrix $\mathbf{A}_{\alpha }^{\
\underline{\alpha }}(u)$ (\ref{4vtn2}), or its generalization to a matrix $%
\widetilde{\mathbf{A}}_{\alpha }^{\ \underline{\alpha }}$ (\ref{4qefp}), can
be not derived only from solutions of (\ref{4eq01}). Such transforms define
certain, in general, nonintegrable distributions related to new classes of
Einstein equations.

\subsection{Two parameter transforms of nonholonomic solutions}

As a matter of principle, any first type parameter transform can be
represented as a generalized anholonomic frame transform labelled by an
additional parameter. It should be also noted that there are two
possibilities to define superpositions of the parameter transforms and
anholonomic frame deformations both resulting in new classes of exact
solutions of the vacuum Einstein equations. In the first case, we start with
a parameter transform and, in the second case, the anholonomic deformations
are considered from the very beginning. The aim of this section is to
examine such possibilities.

Firstly, let us consider an exact vacuum solution $\ _{\shortmid }^{\circ }%
\mathbf{g}$\ (\ref{4ds}) in Einstein gravity generated following the
anholonomic frame method. Even it is generic off--diagonal and depends on
various types of integration functions and constants, it is obvious that it
possess at least a Killing vector symmetry because the metric does not
depend on variable $y^{5}.$ We can apply the first type parameter transform
to a such metric generated by anholonomic deforms (\ref{4nfd}). If we work
in a coordinate base with the coefficients of $\ _{\shortmid }^{\circ }%
\mathbf{g}$ defined in the form $\ _{\shortmid }^{\circ }\underline{g}%
_{\alpha \beta }=\ _{\shortmid }^{\circ }g_{\underline{\alpha }\underline{%
\beta }},$ we generate a set of exact solutions
\begin{equation*}
\ _{\shortmid }^{\circ }\underline{\widetilde{g}}_{\alpha \beta }\mathbf{(}%
\theta ^{\prime }\mathbf{)}=\widetilde{B}_{\alpha }^{\ \alpha ^{\prime
}}(\theta ^{\prime })\ \widetilde{B}_{\alpha }^{\ \beta ^{\prime }}(\theta
^{\prime })\ _{\shortmid }^{\circ }\underline{g}_{\alpha ^{\prime }\beta
^{\prime }},
\end{equation*}%
see (\ref{4ger1t}), were the transforms (\ref{4mgt}), labelled by a
parameter $\theta ^{\prime },$ are not adapted to a nonholonomic $(n+m)$%
--splitting. We can elaborate N--adapted constructions starting with an
exact solution parametrized in the form (\ref{4m2}), for instance, like $\
_{\shortmid }^{\circ }\mathbf{g_{\alpha ^{\prime }\beta ^{\prime }}=A}%
_{\alpha ^{\prime }}^{\ \underline{\alpha }}\mathbf{A}_{\beta ^{\prime }}^{\
\underline{\beta }}\check{g}_{\underline{\alpha }\underline{\beta }}$ , with
$\mathbf{A}_{\alpha }^{\ \underline{\alpha }}$ being of type (\ref{4vtn2})
with coefficients satisfying the conditions (\ref{4polf1}). The target
'boldface' solutions are generated as transforms
\begin{equation}
\ _{\shortmid }^{\circ }\widetilde{\mathbf{g}}_{\alpha \beta }\mathbf{(}%
\theta ^{\prime }\mathbf{)}=\widetilde{\mathbf{B}}_{\alpha }^{\ \alpha
^{\prime }}(\theta ^{\prime })\ \widetilde{\mathbf{B}}_{\alpha }^{\ \beta
^{\prime }}(\theta ^{\prime })\ _{\shortmid }^{\circ }\mathbf{g}_{\alpha
^{\prime }\beta ^{\prime }},  \label{4pts}
\end{equation}%
where
\begin{equation*}
\widetilde{\mathbf{B}}_{\alpha }^{\ \alpha ^{\prime }}=\widetilde{\mathbf{A}}%
_{\alpha }^{\ \underline{\alpha }}\ \ _{\shortmid }^{\circ }\mathbf{A}_{%
\underline{\alpha }}^{\ \alpha ^{\prime }},
\end{equation*}%
like in (\ref{4mgt}), but for ''boldfaced' objects, the matrix $\
_{\shortmid }^{\circ }\mathbf{A}_{\underline{\alpha }}^{\ \alpha ^{\prime }}$
is inverse to
\begin{equation*}
\ _{\shortmid }^{\circ }\mathbf{A}_{\alpha ^{\prime }}^{\ \underline{\alpha }%
}(u)=\left[
\begin{array}{cc}
\sqrt{|\eta _{i^{\prime }}|}\delta _{i^{\prime }}^{\ \underline{i}} & -\eta
_{i^{\prime }}^{b^{\prime }}\check{N}_{j^{\prime }}^{b^{\prime }}\delta
_{b^{\prime }}^{\ \underline{a}} \\
0 & \sqrt{|\eta _{a^{\prime }}|}\delta _{a^{\prime }}^{\ \underline{a}}%
\end{array}%
\right]
\end{equation*}%
and there is considered the matrix
\begin{equation*}
\widetilde{\mathbf{A}}_{\alpha }^{\ \underline{\alpha }}(u,\theta ^{\prime
})=\left[
\begin{array}{cc}
\sqrt{|\eta _{i}\ \widetilde{\eta }_{i}(\theta ^{\prime })|}\delta
_{i^{\prime }}^{\ \underline{i}} & -\eta _{i}^{b}\ \widetilde{\eta }%
_{i}^{b}(\theta ^{\prime })\check{N}_{j}^{b}\delta _{b}^{\ \underline{a}} \\
0 & \sqrt{|\eta _{a}\ \widetilde{\eta }_{a}(\theta ^{\prime })|}\delta
_{a}^{\ \underline{a}}%
\end{array}%
\right] ,
\end{equation*}%
where $\widetilde{\eta }_{i}(u,\theta ^{\prime }),\widetilde{\eta }%
_{a}(u,\theta ^{\prime })$ and $\widetilde{\eta }_{i}^{a}(u,\theta ^{\prime
})$ are gravitational polarizations of type (\ref{4polf}).\footnote{%
we do not summarize on repeating two indices if they both are of lower/
upper type} Here it should be emphasized that even $\ _{\shortmid }^{\circ }%
\widetilde{\mathbf{g}}_{\alpha \beta }(\theta ^{\prime })$ are exact
solutions of the vacuum Einstein equations they can not be represented by
ansatz of type (\ref{4dst}), with $\theta \rightarrow \theta ^{\prime },$
because the mentioned polarizations were not constrained to be of type (\ref%
{4polf1}) and satisfy any conditions of type (\ref{4const6}).\footnote{%
As a matter of principle, we can deform nonholonomically any solution from
the family $\ _{\shortmid }^{\circ }\widetilde{\mathbf{g}}_{\alpha \beta
}(\theta ^{\prime })$ to an ansatz of type (\ref{4dst}).}

Now we prove that by using superpositions of nonholonomic and parameter
transforms we can generate two parameter families of solutions. This is
possible, for instance, if the metric $\ _{\shortmid }^{\circ }\mathbf{g}%
_{\alpha ^{\prime }\beta ^{\prime }}$ form (\ref{4pts}), in its turn, was
generated as an ansatz of type (\ref{4dst}), from another exact solution $\
_{\shortmid }^{\circ }\mathbf{g}_{\alpha ^{\prime \prime }\beta ^{\prime
\prime }}.$ We write
\begin{equation*}
\ _{\shortmid }^{\circ }\mathbf{g}_{\alpha ^{\prime }\beta ^{\prime
}}(\theta )=\widetilde{\mathbf{B}}_{\alpha ^{\prime }}^{\ \alpha ^{\prime
\prime }}(u,\theta )\ \widetilde{\mathbf{B}}_{\beta ^{\prime }}^{\ \beta
^{\prime \prime }}(u,\theta )\ _{\shortmid }^{\circ }\mathbf{g}_{\alpha
^{\prime \prime }\beta ^{\prime \prime }}
\end{equation*}%
and define the superposition of transforms
\begin{equation}
\ _{\shortmid }^{\circ }\widetilde{\mathbf{g}}_{\alpha \beta }\mathbf{(}%
\theta ^{\prime },\theta \mathbf{)}=\widetilde{\mathbf{B}}_{\alpha }^{\
\alpha ^{\prime }}(\theta ^{\prime })\ \widetilde{\mathbf{B}}_{\alpha }^{\
\beta ^{\prime }}(\theta ^{\prime })\ \widetilde{\mathbf{B}}_{\alpha
^{\prime }}^{\ \alpha ^{\prime \prime }}(\theta )\ \widetilde{\mathbf{B}}%
_{\beta ^{\prime }}^{\ \beta ^{\prime \prime }}(\theta )\ _{\shortmid
}^{\circ }\mathbf{g}_{\alpha ^{\prime \prime }\beta ^{\prime \prime }}.
\label{4pts1}
\end{equation}%
It can be considered an iteration procedure of nonholonomic parameter
transforms of type (\ref{4pts1}) when an exact vacuum solution of the
Einstein equations is related via a multi $\theta $--parameters frame map
with another prescribed vacuum solution. Using anholonomic deformations, one
introduces (into chains of such transforms) certain classes of metrics which
are not exact solutions but nonholonomically deformed from, or to, some
exact solutions.

\subsection{Multi--parametric Einstein spaces and Ricci flows}

Let us denote by $\overleftrightarrow{\theta }=\left( ~^{k}\theta =\theta
^{\prime },~^{2}\theta ,...,\theta =~^{1}\theta ,\right) $ a chain of
nonholonomic parametric transforms (it can be more general as (\ref{4pts1}),
beginning with an arbitrary metric $\mathbf{g}_{\alpha ^{\prime \prime
}\beta ^{\prime \prime }})$ resulting in a metric $\ \widetilde{\mathbf{g}}%
_{\alpha \beta }\mathbf{(}\overleftrightarrow{\theta }\mathbf{).}$ Any step
of nonholonomic parametric and/ or frame transforms are parametrize matrices
of type (\ref{4qefp}), (\ref{4mgt1}) or (\ref{4pts}). Here, for simplicity,
we consider two important examples when $\ \widetilde{\mathbf{g}}_{\alpha
\beta }\mathbf{(}\overleftrightarrow{\theta }\mathbf{)}$ will generate
solutions of the nonholonomic Einstein equations or Ricci flow equations.

\subsubsection{Example 1:}

We get a multi--parametric ansatz of type (\ref{4ans5d}) with $h_{4}^{\ast
}\neq 0$ and $h_{5}^{\ast }\neq 0$ if $\widetilde{\mathbf{g}}_{\alpha \beta }%
\mathbf{(}\overleftrightarrow{\theta }\mathbf{)}$ is of type
\begin{eqnarray}
\ ^{\circ }\mathbf{g(}\overleftrightarrow{\theta }\mathbf{)} &=&\epsilon
_{1}\ dx^{1}\otimes dx^{1}+\epsilon _{2}g_{2}(\overleftrightarrow{\theta }%
,x^{\widehat{i}})\ dx^{2}\otimes dx^{2}  \notag \\
&&+\epsilon _{3}g_{3}(\overleftrightarrow{\theta },x^{\widehat{i}})\
dx^{3}\otimes dx^{3}  \notag \\
&&+\epsilon _{4}h_{0}^{2}(\overleftrightarrow{\theta },x^{i})\left[ f^{\ast
}\left( \overleftrightarrow{\theta },x^{i},v\right) \right] ^{2}|\varsigma
\left( \overleftrightarrow{\theta },x^{i},v\right) |\ \delta v\otimes \delta
v  \notag \\
&&+\epsilon _{5}\left[ f\left( \overleftrightarrow{\theta },x^{i},v\right)
-f_{0}(\overleftrightarrow{\theta },x^{i})\right] ^{2}\ \delta y^{5}\otimes
\delta y^{5},  \label{4gensol1m} \\
\delta v &=&dv+w_{k}\left( \overleftrightarrow{\theta },x^{i},v\right)
dx^{k},\ \delta y^{5}= dy^{5}+n_{k}\left(\overleftrightarrow{\theta}%
,x^{i},v\right) dx^{k},  \notag
\end{eqnarray}
the indices with ''hat'' running the values $\widehat{i},\widehat{j},...=2,3$%
, where\ $g_{\widehat{k}}\left( \overleftrightarrow{\theta },x^{\widehat{i}%
}\right) $ are multi--paramet\-ric families of solutions of the 2D equation (%
\ref{4ep1a}) for given sources $\Upsilon _{4}\left( \overleftrightarrow{%
\theta },x^{\widehat{i}}\right) ,$%
\begin{eqnarray*}
&&\varsigma \left( \overleftrightarrow{\theta },x^{i},v\right) =\varsigma
_{\lbrack 0]}\left( \overleftrightarrow{\theta },x^{i}\right) -\frac{%
\epsilon _{4}}{8}h_{0}^{2}(\overleftrightarrow{\theta },x^{i})\times \\
&&\int \Upsilon _{2}(\overleftrightarrow{\theta },x^{\widehat{k}},v)f^{\ast
}\left( \overleftrightarrow{\theta },x^{i},v\right) \left[ f\left(
\overleftrightarrow{\theta },x^{i},v\right) -f_{0}(\overleftrightarrow{%
\theta },x^{i})\right] dv,
\end{eqnarray*}%
and the N--connection $N_{i}^{4}=w_{i}(\overleftrightarrow{\theta }%
,x^{k},v), $ $N_{i}^{5}=n_{i}(\overleftrightarrow{\theta },x^{k},v) $
computed
\begin{eqnarray}
&& w_{i}\left( \overleftrightarrow{\theta },x^{k},v\right) =-\frac{\partial
_{i}\varsigma \left( \overleftrightarrow{\theta },x^{k},v\right) }{\varsigma
^{\ast }\left( \overleftrightarrow{\theta },x^{k},v\right) },
\label{4gensol1wm} \\
&&n_{k}\left( \overleftrightarrow{\theta },x^{k},v\right) =n_{k[1]}\left(
\overleftrightarrow{\theta },x^{i}\right) +n_{k[2]}\left(
\overleftrightarrow{\theta },x^{i}\right) \times  \label{4gensol1nm} \\
&&\int \frac{\left[ f^{\ast }\left( \overleftrightarrow{\theta }%
,x^{i},v\right) \right] ^{2}}{\left[ f\left( \overleftrightarrow{\theta }%
,x^{i},v\right) -f_{0}(x^{i})\right] ^{3}}\varsigma \left(
\overleftrightarrow{\theta },x^{i},v\right) dv,  \notag
\end{eqnarray}%
define an exact solution of the Einstein equations (\ref{4ep1b}). We
emphasize that such solutions depend on an arbitrary nontrivial function $%
f\left( \overleftrightarrow{\theta },x^{i},v\right),$ with $f^{\ast }\neq 0,$
integration functions $f_{0}(\overleftrightarrow{\theta },x^{i}),$ $%
h_{0}^{2}(\overleftrightarrow{\theta },x^{i})$, $\ \varsigma _{\lbrack
0]}\left( \overleftrightarrow{\theta },x^{i}\right) ,$ $n_{k[1]}\left(
\overleftrightarrow{\theta },x^{i}\right) ,$ $n_{k[2]}\left(
\overleftrightarrow{\theta },x^{i}\right) $ and sources $\Upsilon _{2}(%
\overleftrightarrow{\theta },x^{\widehat{k}},v),\Upsilon _{4}\left(
\overleftrightarrow{\theta },x^{\widehat{i}}\right).$ Such values for the
corresponding signatures $\epsilon _{\alpha }=\pm 1$ have to be defined by
certain boundary conditions and physical considerations. We note that
formulas (\ref{4gensol1wm}) and (\ref{4gensol1nm}) state symbolically that
at any intermediary step from the chain $\overleftrightarrow{\theta }$ one
construct the solution following the respective formulas (\ref{4gensol1w})
and (\ref{4gensol1n}). The final aim, is to get a set of metrics (\ref%
{4gensol1m}), parametrized by $\overleftrightarrow{\theta },$ when for fixed
values of $\theta $--parameters, we get solutions of type (\ref{4gensol1}),
for the vacuum Einstein equations for the canonical d--connection.

\subsubsection{Example 2:}

We consider a family of ansatz, labelled by a set of parameters $%
\overleftrightarrow{\theta }$ and $\chi $ (as a matter of principle, we can
identify the Ricci flow parameter $\chi $ with any $\theta $ from the set $%
\overleftrightarrow{\theta }$ considering that the evolution parameter is
also related to the invariance of Killing equations, see Appendix \ref{4apkv}%
),
\begin{eqnarray}
\ _{\shortmid }^{\lambda }\mathbf{g}(\overleftrightarrow{\theta },\chi )
&=&e^{\psi (\overleftrightarrow{\theta },x^{2},x^{3},\chi )}\left[ \epsilon
_{2}\ dx^{2}\otimes dx^{2}+\epsilon _{3}\ dx^{3}\otimes dx^{3}\right] +
\label{4slerf1et} \\
&&h_{4}\left( \overleftrightarrow{\theta },x^{2},x^{3},v,\chi \right) \
\delta v\otimes \delta v  \notag \\
&&+h_{5}\left( \overleftrightarrow{\theta },x^{2},x^{3},v,\chi \right) \
^{\chi }\delta y^{5}\otimes \ ^{\chi }\delta y^{5},  \notag \\
\delta v &=&dv+w_{2}\left( \overleftrightarrow{\theta },x^{2},x^{3},v\right)
dx^{2}+w_{3}\left( \overleftrightarrow{\theta },x^{2},x^{3},v\right) dx^{3},
\notag \\
\ \ ^{\chi }\delta y^{5} &=&dy^{5}+n_{2}\left( \overleftrightarrow{\theta }%
,x^{2},x^{3},\chi \right) [dx^{2}+dx^{3}],  \notag
\end{eqnarray}%
which for any fixed set $\overleftrightarrow{\theta }$ is of type (\ref%
{4slerf1e}) with the coefficients are subjected to the conditions (\ref%
{4ep2be}), in our case generalized in the form
\begin{eqnarray}
\epsilon _{2}\psi ^{\bullet \bullet }(\overleftrightarrow{\theta },\chi
)+\epsilon _{3}\psi ^{^{\prime \prime }}(\overleftrightarrow{\theta },\chi )
&=&\lambda ,  \notag \\
h_{5}^{\ast }(\overleftrightarrow{\theta })\phi (\overleftrightarrow{\theta }%
)/h_{4}(\overleftrightarrow{\theta })h_{5}(\overleftrightarrow{\theta })
&=&\lambda ,  \label{4ep2bet} \\
w_{2}^{\prime }(\overleftrightarrow{\theta })-w_{3}^{\bullet }(%
\overleftrightarrow{\theta })+w_{3}(\overleftrightarrow{\theta })w_{2}^{\ast
}(\overleftrightarrow{\theta })-w_{2}(\overleftrightarrow{\theta }%
)w_{3}^{\ast }(\overleftrightarrow{\theta }) &=&0,  \notag \\
n_{2}^{\prime }(\overleftrightarrow{\theta },\chi )-n_{2}^{\bullet }(%
\overleftrightarrow{\theta },\chi ) &=&0,  \notag
\end{eqnarray}%
for $w_{\widehat{i}}=\partial _{\widehat{i}}\phi /\phi ^{\ast },$ see (\ref%
{4coefa}), being compatible with (\ref{4cond3a}) and (\ref{4cond3b}) and
considered that finally on solve the Einstein equations for given surces $%
\Upsilon _{4}=\lambda $ and $\Upsilon _{2}=\lambda .$ The metrics (\ref%
{4slerf1et}) define self--consistent evolutions of a multi--parametric class
of general solutions of the Ricci flow equations (\ref{4eq1})--(\ref{4eq3})
transformed equivalently in the form (\ref{4ep1a})--(\ref{4ep4a}). The
additional constraints (\ref{4ep2bet}) define multi--parametric integral
subvarieties (foliations) when the evolutions are selected for the Levi
Civita connections.

\section{Summary and Discussion}

In this work, we have developed an unified geometric approach to
constructing exact solutions in gravity and Ricci flow theories following
superpositions of anholonomic frame deformations and multi--parametric
transforms with Killing symmetries.

The anholonomic frame method, proposed for generalized Finsler and
Lagran\-ge theories and restricted to the Einstein and string gravity,
applies the formalism of nonholonomic frame deformations \cite%
{4vhep,4vt,4vs,4vesnc} (see outline of results in \cite{4vsgg} and
references therein) when the gravitational field equations transform into
systems of nonlinear partial differential equations which can be integrated
in general form. The new classes of solutions are defined by generic
off--diagonal metrics depending on integration functions on one, two and
three/ four variables (if we consider four or five dimensional, in brief, 5D
or 4D, spacetimes). The important property of such  solutions is that they
can be generalized for effective cosmological constants induced by certain
locally anisotropic matter field interactions, quantum fluctuations and/or
string corrections and from Ricci flow theory.

In general relativity, there is also a method elaborated in Refs. \cite%
{4geroch1,4geroch2} as a general scheme when one (two) parameter families of
exact solutions are defined by any source--free solutions of Einstein's
equations with one (two) Killing vector field(s) (for nonholonomic
manifolds, we call such transforms to be one-, two- or multi--parameter
nonholonomic deformations/ transforms). A successive iteration procedure
results in a class of solutions characterized by an infinite number of
parameters for a non--Abelian group involving arbitrary functions on one
variable.

Both the parametric deformation techniques combined  with nonholonomic
transforms state a number of possibilities to construct ''target'' families
of exact solutions and evolution scenarios starting with primary metrics not
subjected to the conditions to solve the Einstein equations. The new classes
of solutions depend on group like and flow parameters and on sets of
integration functions and constants resulting from the procedure of
integrating systems of partial differential equations to which the field
equations are reduced for certain off--diagonal metric ansatz and
generalized connections. Constraining the integral varieties, for a
corresponding subset of integration functions, the target solutions are
determined to define Einstein spacetimes and their Ricci flow evolutions. In
general, such configurations are nonholonomic but can constrained to define
geometric evolutions for the Levi Civita connections.

\vskip3pt \textbf{Acknowledgement: } The work is performed during a visit at
Fields Institute.

\appendix

\setcounter{equation}{0} \renewcommand{\theequation}
{A.\arabic{equation}} \setcounter{subsection}{0}
\renewcommand{\thesubsection}
{A.\arabic{subsection}}

\section{The Anholonomic Frame Method}

\label{4sah} We outline the geometry of nonholonomic frame deformations and
nonlinear connection (N--connection) structures \cite{4vesnc,4vsgg}.

Let us consider a $(n+m)$--dimensional manifold $\mathbf{V}$\ enabled with a
prescribed frame structure (\ref{4ft}) when frame transforms are linear on $%
N_{i}^{b}(u),$
\begin{eqnarray}
\mathbf{A}_{\alpha }^{\ \underline{\alpha }}(u) &=&\left[
\begin{array}{cc}
e_{i}^{\ \underline{i}}(u) & -N_{i}^{b}(u)e_{b}^{\ \underline{a}}(u) \\
0 & e_{a}^{\ \underline{a}}(u)%
\end{array}%
\right] ,  \label{4vt1} \\
\mathbf{A}_{\ \underline{\beta }}^{\beta }(u) &=&\left[
\begin{array}{cc}
e_{\ \underline{i}}^{i\ }(u) & N_{k}^{b}(u)e_{\ \underline{i}}^{k\ }(u) \\
0 & e_{\ \underline{a}}^{a\ }(u)%
\end{array}%
\right] ,  \label{4vt2}
\end{eqnarray}%
where $i,j,..=1,2,...,n$ and $a,b,...=n+1,n+2,...n+m$ and $u=\{u^{\alpha
}=(x^{i},y^{a})\}$ are local coordinates. The geometric constructions will
be adapted to a conventional $n+m$ splitting stated by a set of coefficients
$\mathbf{N}=\{N_{i}^{a}(u)\}$ defining a nonlinear connection
(N--connection) structure as a nonintergrable distribution%
\begin{equation}
T\mathbf{V=}h\mathbf{V\oplus }v\mathbf{V}  \label{4distr}
\end{equation}%
with a conventional horizontal (h) subspace, $h\mathbf{V,}$ (with geometric
objects labelled by ''horizontal'' indices $i,j,...)$ and vertical (v)
subspace $v\mathbf{V}$ (with geometric objects labelled by indices $%
a,b,...). $ The ''boldfaced'' symbols will be used to emphasize that certain
spaces (geometric objects) are provided (adapted) with (to) a N--connection
structure $\mathbf{N.}$

The transforms (\ref{4vt1}) and (\ref{4vt2}) define a N--adapted frame
structure
\begin{equation}
\mathbf{e}_{\nu }=\left( \mathbf{e}_{i}=\partial _{i}-N_{i}^{a}(u)\partial
_{a},e_{a}=\partial _{a}\right) ,  \label{4dder}
\end{equation}%
and the dual frame (coframe) structure%
\begin{equation}
\mathbf{e}^{\mu }=\left( e^{i}=dx^{i},\mathbf{e}%
^{a}=dy^{a}+N_{i}^{a}(u)dx^{i}\right) .  \label{4ddif}
\end{equation}%
The frames (\ref{4ddif}) satisfy the certain nonholonomy (equivalently,
anholonomy) relations of type (\ref{4anhr}),
\begin{equation}
\lbrack \mathbf{e}_{\alpha },\mathbf{e}_{\beta }]=\mathbf{e}_{\alpha }%
\mathbf{e}_{\beta }-\mathbf{e}_{\beta }\mathbf{e}_{\alpha }=W_{\alpha \beta
}^{\gamma }\mathbf{e}_{\gamma },  \label{4nanhrel}
\end{equation}%
with anholonomy coefficients
\begin{equation}
W_{ia}^{b}=\partial _{a}N_{i}^{b}\mbox{ and }W_{ji}^{a}=\Omega _{ij}^{a}=%
\mathbf{e}_{j}(N_{i}^{a})-\mathbf{e}_{j}(N_{i}^{a}).  \label{4anhncc}
\end{equation}%
A distribution (\ref{4distr}) is integrable, i.e. $\mathbf{V}$ is a
foliation, if and only if the coefficients defined by $\mathbf{N}%
=\{N_{i}^{a}(u)\}$ satisfy the condition $\Omega _{ij}^{a}=0.$ In general, a
spacetime with prescribed nonholonomic splitting into h- and v--subspaces
can be considered as a nonholonomic manifold \cite{4vesnc,4bejf,4esv}.

Let us consider a metric structure on $\mathbf{V},$%
\begin{equation}
\ \breve{g}=\underline{g}_{\alpha \beta }\left( u\right) du^{\alpha }\otimes
du^{\beta }  \label{4metr}
\end{equation}%
defined by coefficients%
\begin{equation}
\underline{g}_{\alpha \beta }=\left[
\begin{array}{cc}
g_{ij}+N_{i}^{a}N_{j}^{b}h_{ab} & N_{j}^{e}h_{ae} \\
N_{i}^{e}h_{be} & h_{ab}%
\end{array}%
\right] .  \label{4ansatz}
\end{equation}%
This metric\ is generic off--diagonal, i.e. it can not be diagonalized by
any coordinate transforms if $N_{i}^{a}(u)$ are any general functions. We
can adapt the metric (\ref{4metr}) to a N--connection structure $\mathbf{N}%
=\{N_{i}^{a}(u)\}$ induced by the off--diagonal coefficients in (\ref%
{4ansatz}) if we impose that the conditions
\begin{equation*}
\breve{g}(e_{i},\ e_{a})=0,\mbox{ equivalently, }\underline{g}%
_{ia}-N_{i}^{b}h_{ab}=0,
\end{equation*}%
where $\underline{g}_{ia}$ $\doteqdot g(\partial /\partial x^{i},\partial
/\partial y^{a}),$ are satisfied for the corresponding local basis (\ref%
{4dder}). In this case $N_{i}^{b}=h^{ab}\underline{g}_{ia},$ where $h^{ab}$
is inverse to $h_{ab},$ and we can write the metric $\breve{g}$ (\ref%
{4ansatz})\ in equivalent form, as a distinguished metric (d--metric)
adapted to a N--connection structure,
\begin{equation}
\mathbf{g}=\mathbf{g}_{\alpha \beta }\left( u\right) \mathbf{e}^{\alpha
}\otimes \mathbf{e}^{\beta }=g_{ij}\left( u\right) e^{i}\otimes
e^{j}+h_{ab}\left( u\right) \ \mathbf{e}^{a}\otimes \ \mathbf{e}^{b},
\label{4dmetr}
\end{equation}%
where $g_{ij}\doteqdot \mathbf{g}\left( \mathbf{e}_{i},\mathbf{e}_{j}\right)
$ and $h_{ab}\doteqdot \mathbf{g}\left( e_{a},e_{b}\right) .$ The
coefficients $\mathbf{g}_{\alpha \beta }$ and $\underline{g}_{\alpha \beta
}=g_{\underline{\alpha }\underline{\beta }}$ are related by formulas%
\begin{equation}
\mathbf{g}_{\alpha \beta }=\mathbf{A}_{\alpha }^{\ \underline{\alpha }}%
\mathbf{A}_{\beta }^{\ \underline{\beta }}g_{\underline{\alpha }\underline{%
\beta }},  \label{4fmt}
\end{equation}%
or
\begin{equation*}
g_{ij}=e_{i}^{\ \underline{i}}e_{j}^{\ \underline{j}}g_{\underline{i}%
\underline{j}}\mbox{ and }h_{ab}=e_{a}^{\ \underline{a}}e_{b}^{\ \underline{b%
}}g_{\underline{a}\underline{b}},
\end{equation*}%
where the frame transform is given by matrices (\ref{4vt1}) with $e_{i}^{\
\underline{i}}=\delta _{i}^{\ \underline{i}}$ and $e_{a}^{\ \underline{a}%
}=\delta _{a}^{\ \underline{a}}.$ We shall call some geometric objects, for
instance, tensors, connections,..., to be distinguished by a N--connection
structure, in brief, d--tensors, d--connections,... if they are stated by
components computed with respect to N--adapted frames (\ref{4dder}) and (\ref%
{4ddif}). In this case, the geometric constructions are elaborated in
N--adapted form, i.e. they are adapted to the nonholonomic distribution (\ref%
{4distr}).

Any vector field $\mathbf{X=(}hX\mathbf{,}\ vX\mathbf{)}$ on $T\mathbf{V}$
can be written in N--adapted form as a d--vector%
\begin{equation*}
\mathbf{X=}X^{\alpha }\mathbf{e}_{\alpha }=\mathbf{(}hX=X^{i}\mathbf{e}_{i}%
\mathbf{,\ }vX=X^{a}e_{a}\mathbf{).}
\end{equation*}%
In a similar form, we can 'N--adapt' any tensor object and get a d--tensor.

By definition, a d--connection is adapted to the distribution (\ref{4distr})
and splits into h-- and v--covariant derivatives, $\mathbf{D}=hD+\ vD,$
where $hD=\{\mathbf{D}_{k}=\left( L_{jk}^{i},L_{bk\;}^{a}\right) \}$ and $\
vD=\{\mathbf{D}_{c}=\left( C_{jk}^{i},C_{bc}^{a}\right) \}$ are
correspondingly introduced as h- and v--parametrizations of the coefficients%
\begin{equation*}
L_{jk}^{i}=\left( \mathbf{D}_{k}\mathbf{e}_{j}\right) \rfloor e^{i},\quad
L_{bk}^{a}=\left( \mathbf{D}_{k}e_{b}\right) \rfloor \mathbf{e}%
^{a},~C_{jc}^{i}=\left( \mathbf{D}_{c}\mathbf{e}_{j}\right) \rfloor
e^{i},\quad C_{bc}^{a}=\left( \mathbf{D}_{c}e_{b}\right) \rfloor \mathbf{e}%
^{a}.
\end{equation*}%
The components $\mathbf{\Gamma }_{\ \alpha \beta }^{\gamma }=\left(
L_{jk}^{i},L_{bk}^{a},C_{jc}^{i},C_{bc}^{a}\right) ,$ with the coefficients
defined with respect to (\ref{4ddif}) and (\ref{4dder}), completely define a
d--connection $\mathbf{D}$ on a N--anholonomic manifold $\mathbf{V}.$

The simplest way to perform a local covariant calculus by applying
d--connecti\-ons is to use N--adapted differential forms and to introduce
the d--connection 1--form $\mathbf{\Gamma }_{\ \beta }^{\alpha }=\mathbf{%
\Gamma }_{\ \beta \gamma }^{\alpha }\mathbf{e}^{\gamma },$ when the
N--adapted components of d-connection $\mathbf{D}_{\alpha }=(\mathbf{e}%
_{\alpha }\rfloor \mathbf{D})$ are computed following formulas
\begin{equation}
\mathbf{\Gamma }_{\ \alpha \beta }^{\gamma }\left( u\right) =\left( \mathbf{D%
}_{\alpha }\mathbf{e}_{\beta }\right) \rfloor \mathbf{e}^{\gamma },
\label{4cond2}
\end{equation}%
where ''$\rfloor "$ denotes the interior product. We define in N--adapted
form the torsion $\mathbf{T=\{\mathcal{T}^{\alpha }\}}$ (\ref{4tors}),
\begin{equation}
\mathcal{T}^{\alpha }\doteqdot \mathbf{De}^{\alpha }=d\mathbf{e}^{\alpha }+%
\mathbf{\Gamma }_{\ \beta }^{\alpha }\wedge \mathbf{e}_{\alpha },
\label{4torsa}
\end{equation}%
and curvature $\mathbf{R}=\{\mathcal{R}_{\ \beta }^{\alpha }\}$ (\ref{4curv}%
),
\begin{equation}
\mathcal{R}_{\ \beta }^{\alpha }\doteqdot \mathbf{D\Gamma }_{\beta }^{\alpha
}=d\mathbf{\Gamma }_{\beta }^{\alpha }-\mathbf{\Gamma }_{\ \beta }^{\gamma
}\wedge \mathbf{\Gamma }_{\ \gamma }^{\alpha }.  \label{4curva}
\end{equation}

The coefficients of torsion $\mathbf{T}$ (\ref{4torsa}) of a d--connection $%
\mathbf{D}$ (in\ brief, d--torsion) are computed with respect to N--adapted
frames (\ref{4ddif}) and (\ref{4dder}),
\begin{eqnarray}
T_{\ jk}^{i} &=&L_{\ jk}^{i}-L_{\ kj}^{i},\ T_{\ ja}^{i}=-T_{\ aj}^{i}=C_{\
ja}^{i},\ T_{\ ji}^{a}=\Omega _{\ ji}^{a},\   \notag \\
T_{\ bi}^{a} &=&T_{\ ib}^{a}=\frac{\partial N_{i}^{a}}{\partial y^{b}}-L_{\
bi}^{a},\ T_{\ bc}^{a}=C_{\ bc}^{a}-C_{\ cb}^{a},  \label{4dtors}
\end{eqnarray}%
where, for instance, $T_{\ jk}^{i}$ and $T_{\ bc}^{a}$ are respectively the
coefficients of the $h(hh)$--torsion $hT(hX,hY)$ and $v(vv)$--torsion $%
\mathbf{\ }vT(\mathbf{\ }vX,\mathbf{\ }vY).$ In a similar form, we can
compute the coefficients of a curvature $\mathbf{R,}$ d--curvatures.

There is a preferred, canonical d--connection structure,$\ \widehat{\mathbf{D%
}}\mathbf{,}$ $\ $on a N--anholonomic manifold $\mathbf{V}$ constructed only
from the metric and N--con\-nec\-ti\-on coefficients $%
[g_{ij},h_{ab},N_{i}^{a}]$ and satisfying the conditions $\widehat{\mathbf{D}%
}\mathbf{g}=0$ and $\widehat{T}_{\ jk}^{i}=0$ and $\widehat{T}_{\ bc}^{a}=0.$
It should be noted that, in general, the components $\widehat{T}_{\
ja}^{i},\ \widehat{T}_{\ ji}^{a}$ and $\widehat{T}_{\ bi}^{a}$ are not zero.
This is an anholonomic frame (equivalently, off--diagonal metric) effect.
Hereafter, we consider only geometric constructions with the canonical
d--connection which allow, for simplicity, to omit ''hats'' on d--objects.
We can verify by straightforward calculations that the linear connection $%
\mathbf{\Gamma }_{\ \alpha \beta }^{\gamma }=\left(
L_{jk}^{i},L_{bk}^{a},C_{jc}^{i},C_{bc}^{a}\right) $ with the coefficients
defined
\begin{equation*}
\mathbf{D}_{\mathbf{e}_{k}}(\mathbf{e}_{j})=L_{jk}^{i}\mathbf{e}_{i},\
\mathbf{D}_{\mathbf{e}_{k}}(e_{b})=L_{bk}^{a}e_{a},\ \mathbf{D}_{e_{b}}(%
\mathbf{e}_{j})=C_{jb}^{i}\mathbf{e}_{i},\ \mathbf{D}%
_{e_{c}}(e_{b})=C_{bc}^{a}e_{a},
\end{equation*}%
where
\begin{eqnarray}
L_{jk}^{i} &=&\frac{1}{2}g^{ir}\left( \mathbf{e}_{k}g_{jr}+\mathbf{e}%
_{j}g_{kr}-\mathbf{e}_{r}g_{jk}\right) ,  \notag \\
L_{bk}^{a} &=&e_{b}(N_{k}^{a})+\frac{1}{2}h^{ac}\left( \mathbf{e}%
_{k}h_{bc}-h_{dc}\ e_{b}N_{k}^{d}-h_{db}\ e_{c}N_{k}^{d}\right) ,
\label{4candcon} \\
C_{jc}^{i} &=&\frac{1}{2}g^{ik}e_{c}g_{jk},\ C_{bc}^{a}=\frac{1}{2}%
h^{ad}\left( e_{c}h_{bd}+e_{c}h_{cd}-e_{d}h_{bc}\right) ,  \notag
\end{eqnarray}%
uniquely solve the conditions stated for the canonical d--connection.

The Levi Civita linear connection $\bigtriangledown =\{\ _{\shortmid }\Gamma
_{\beta \gamma }^{\alpha }\},$ uniquely defined by the conditions $^{\nabla
}T=0$ and $\bigtriangledown \breve{g}=0,$ is not adapted to the distribution
(\ref{4distr}). Denoting $_{\shortmid }\Gamma _{\beta \gamma }^{\alpha }=(\
_{\shortmid }L_{jk}^{i},\ _{\shortmid }L_{jk}^{a},\ _{\shortmid
}L_{bk}^{i},\ _{\shortmid }L_{bk}^{a},\ _{\shortmid }C_{jb}^{i},\
_{\shortmid }C_{jb}^{a},\ _{\shortmid }C_{bc}^{i},\ _{\shortmid }C_{bc}^{a}),
$ for
\begin{eqnarray*}
\bigtriangledown _{\mathbf{e}_{k}}(\mathbf{e}_{j}) &=&\ _{\shortmid
}L_{jk}^{i}\mathbf{e}_{i}+\ _{\shortmid }L_{jk}^{a}e_{a},\ \bigtriangledown
_{\mathbf{e}_{k}}(e_{b})=\ _{\shortmid }L_{bk}^{i}\mathbf{e}_{i}+\
_{\shortmid }L_{bk}^{a}e_{a}, \\
\bigtriangledown _{e_{b}}(\mathbf{e}_{j}) &=&\ _{\shortmid }C_{jb}^{i}%
\mathbf{e}_{i}+\ _{\shortmid }C_{jb}^{a}e_{a},\ \bigtriangledown
_{e_{c}}(e_{b})=\ _{\shortmid }C_{bc}^{i}\mathbf{e}_{i}+\ _{\shortmid
}C_{bc}^{a}e_{a},
\end{eqnarray*}%
after a straightforward calculus we get
\begin{eqnarray}
\ _{\shortmid }L_{jk}^{i} &=&L_{jk}^{i},\ _{\shortmid
}L_{jk}^{a}=-C_{jb}^{i}g_{ik}h^{ab}-\frac{1}{2}\Omega _{jk}^{a},
\label{4lccon} \\
\ _{\shortmid }L_{bk}^{i} &=&\frac{1}{2}\Omega _{jk}^{c}h_{cb}g^{ji}-\frac{1%
}{2}(\delta _{j}^{i}\delta _{k}^{h}-g_{jk}g^{ih})C_{hb}^{j},  \notag \\
\ _{\shortmid }L_{bk}^{a} &=&L_{bk}^{a}+\frac{1}{2}(\delta _{c}^{a}\delta
_{d}^{b}+h_{cd}h^{ab})\left[ L_{bk}^{c}-e_{b}(N_{k}^{c})\right] ,  \notag \\
\ _{\shortmid }C_{kb}^{i} &=&C_{kb}^{i}+\frac{1}{2}\Omega
_{jk}^{a}h_{cb}g^{ji}+\frac{1}{2}(\delta _{j}^{i}\delta
_{k}^{h}-g_{jk}g^{ih})C_{hb}^{j},  \notag \\
\ _{\shortmid }C_{jb}^{a} &=&-\frac{1}{2}(\delta _{c}^{a}\delta
_{b}^{d}-h_{cb}h^{ad})\left[ L_{dj}^{c}-e_{d}(N_{j}^{c})\right] ,\
_{\shortmid }C_{bc}^{a}=C_{bc}^{a},  \notag \\
\ _{\shortmid }C_{ab}^{i} &=&-\frac{g^{ij}}{2}\left\{ \left[
L_{aj}^{c}-e_{a}(N_{j}^{c})\right] h_{cb}+\left[ L_{bj}^{c}-e_{b}(N_{j}^{c})%
\right] h_{ca}\right\} ,  \notag
\end{eqnarray}%
where $\Omega _{jk}^{a}$ are computed as in the second formula in (\ref%
{4anhncc}).

For our purposes, it is important to state the conditions when both the Levi
Civita connection and the canonical d--connection may be defined by the same
set of coefficients with respect to a fixed frame of reference. Following
formulas (\ref{4candcon}) and (\ref{4lccon}), we obtain equality $%
_{\shortmid }\Gamma _{\beta \gamma }^{\alpha }=\mathbf{\Gamma }_{\ \alpha
\beta }^{\gamma }$ if%
\begin{equation}
\Omega _{jk}^{c}=0  \label{4fols}
\end{equation}%
(there are satisfied the integrability conditions and our manifold admits a
foliation structure),
\begin{equation}
\ _{\shortmid }C_{kb}^{i}=C_{kb}^{i}=0  \label{4coef0}
\end{equation}%
and $L_{aj}^{c}-e_{a}(N_{j}^{c})=0,$ which, following the second formula in (%
\ref{4candcon}), is equivalent to%
\begin{equation}
\mathbf{e}_{k}h_{bc}-h_{dc}\ e_{b}N_{k}^{d}-h_{db}\ e_{c}N_{k}^{d}=0.
\label{4cond3}
\end{equation}

We conclude this section with the remark that if the conditions (\ref{4fols}%
), (\ref{4coef0}) and (\ref{4cond3}) hold true for the metric (\ref{4metr}),
equivalently (\ref{4dmetr}), the torsion coefficients (\ref{4dtors}) vanish.
This results in respective equalities of the coefficients of the Riemann,
Ricci and Einstein tensors (the conditions (\ref{4cond1}) being satisfied)
for two different linear connections.

\setcounter{equation}{0} \renewcommand{\theequation}
{B.\arabic{equation}} \setcounter{subsection}{0}
\renewcommand{\thesubsection}
{B.\arabic{subsection}}

\section{The Killing Vectors Formalism}

\label{4apkv}

The first parametric method (on holonomic (pseudo) Riemannian spaces, it  is
also called the Geroch method \cite{4geroch1}) proposes a scheme of
constructing a one--para\-met\-er family of vacuum exact solutions (labelled
by tilde $" \widetilde{} "$ and depending on a real parameter $\theta $)
\begin{equation}
\ _{\shortmid }^{\circ }\widetilde{g}(\theta )=\ _{\shortmid }^{\circ }%
\widetilde{g}_{\alpha \beta }\ e^{\alpha }\otimes e^{\beta }  \label{4germ1}
\end{equation}%
beginning with any source--free solution $\ \ _{\shortmid }^{\circ }g=\{\
_{\shortmid }^{\circ }g_{\alpha \beta }\}$ with Killing vector $\xi =\{\xi
_{\alpha }\}$ symmetry satisfying the conditions $\ _{\shortmid }\mathcal{E}%
=0$ (Einstein equations) and $\nabla _{\xi }(\ _{\shortmid }^{\circ }g)=0$
(Killing equations). We denote this 'primary' spacetime $(V,\ _{\shortmid
}^{\circ }g,\xi _{\alpha })$ and follow the conventions: The class of
metrics $\ _{\shortmid }^{\circ }\widetilde{g}$ is generated by the
transforms
\begin{equation}
\ _{\shortmid }^{\circ }\widetilde{g}_{\alpha \beta }=\widetilde{B}_{\alpha
}^{\ \alpha ^{\prime }}(u,\theta )\ \widetilde{B}_{\beta }^{\ \beta ^{\prime
}}(u,\theta )\ _{\shortmid }^{\circ }g_{\alpha ^{\prime }\beta ^{\prime }}
\label{4ger1t}
\end{equation}%
where the matrix $\widetilde{B}_{\alpha }^{\ \alpha ^{\prime }}$ is
parametrized in the form when
\begin{equation}
\ _{\shortmid }^{\circ }\widetilde{g}_{\alpha \beta }=\lambda \widetilde{%
\lambda }^{-1}(\ _{\shortmid }^{\circ }g_{\alpha \beta }-\lambda ^{-1}\xi
_{\alpha }\xi _{\beta })+\widetilde{\lambda }\mu _{\alpha }\mu _{\beta }
\label{4germ1c}
\end{equation}%
for
\begin{eqnarray*}
\widetilde{\lambda } &=&\lambda \lbrack (\cos \theta -\omega \sin \theta
)^{2}+\lambda ^{2}\sin ^{2}\theta ]^{-1} \\
\mu _{\tau } &=&\widetilde{\lambda }^{-1}\xi _{\tau }+\alpha _{\tau }\sin
2\theta -\beta _{\tau }\sin ^{2}\theta .
\end{eqnarray*}%
A rigorous proof \cite{4geroch1} states that the metrics (\ref{4germ1})
define also exact vacuum solutions with$\ _{\shortmid }\widetilde{\mathcal{E}%
}=0$ if and only if the values $\xi _{\alpha },\alpha _{\tau },$ $\mu _{\tau
}$ from (\ref{4germ1c}), subjected to the conditions $\lambda =\xi _{\alpha
}\xi _{\beta }\ _{\shortmid }^{\circ }g^{\alpha \beta },\ \omega =\xi
^{\gamma }\alpha _{\gamma },\xi ^{\gamma }\mu _{\gamma }=\lambda ^{2}+\omega
^{2}-1,$ solve the equations%
\begin{eqnarray}
\nabla _{\alpha }\omega &=&\epsilon _{\alpha \beta \gamma \tau }\xi ^{\beta
}\ \nabla ^{\gamma }\xi ^{\tau },\ \nabla _{\lbrack \alpha }\alpha _{\beta
]} =\frac{1}{2}\epsilon _{\alpha \beta \gamma \tau }\ \nabla ^{\gamma }\xi
^{\tau },  \label{4eq01} \\
\nabla _{\lbrack \alpha }\mu _{\beta ]} &=&2\lambda \ \nabla _{\alpha }\xi
_{\beta }+\omega \epsilon _{\alpha \beta \gamma \tau }\ \nabla ^{\gamma }\xi
^{\tau },  \notag
\end{eqnarray}%
where the Levi Civita connection $\nabla $ is defined by $\ _{\shortmid
}^{\circ }g$ and $\epsilon _{\alpha \beta \gamma \tau }$ is the absolutely
antisymmetric tensor. The existence of solutions for (\ref{4eq01}) (Geroch's
equations) is guaranteed by the Einstein's and Killing equations.

The first type of parametric transforms (\ref{4ger1t}) can be parametrized
by a matrix $\widetilde{B}_{\alpha }^{\ \alpha ^{\prime }}$ with the
coefficients depending functionally on solutions for (\ref{4eq01}). Fixing a
signature $g_{\underline{\alpha }\underline{\beta }}=diag[\pm 1,\pm
1,....\pm 1]$ and a local coordinate system on $(V,\ _{\shortmid }^{\circ
}g,\xi _{\alpha }),$  we define a local frame of reference $e_{\alpha
^{\prime }}=A_{\alpha ^{\prime }}^{\ \underline{\alpha }}(u)\partial _{%
\underline{\alpha }},$ like in (\ref{4ft}), for which
\begin{equation}
_{\shortmid }^{\circ }g_{\alpha ^{\prime }\beta ^{\prime }}=A_{\alpha
^{\prime }}^{\ \underline{\alpha }}A_{\beta ^{\prime }}^{\ \underline{\beta }%
}g_{\underline{\alpha }\underline{\beta }}.  \label{4qe}
\end{equation}%
We note that $A_{\alpha ^{\prime }}^{\ \underline{\alpha }}$ have to be
constructed as a solution of a system of quadratic algebraic equations (\ref%
{4qe}) for given values $g_{\underline{\alpha }\underline{\beta }}$ and $%
_{\shortmid }^{\circ }g_{\alpha ^{\prime }\beta ^{\prime }}.$ In a similar
form, we can write $\widetilde{e}_{\alpha }=\widetilde{A}_{\alpha }^{\
\underline{\alpha }}(\theta ,u)\partial _{\underline{\alpha }}$  and
\begin{equation}
_{\shortmid }^{\circ }\widetilde{g}_{\alpha \beta }=\widetilde{A}_{\alpha
}^{\ \underline{\alpha }}\widetilde{A}_{\beta }^{\ \underline{\beta }}g_{%
\underline{\alpha }\underline{\beta }}.  \label{4qef}
\end{equation}%
The method guarantees that the family of spacetimes $(V,\ _{\shortmid
}^{\circ }\widetilde{g})$ is also vacuum Einstein but for the corresponding
families of Levi Civita connections $\widetilde{\nabla }.$ In explicit form,
the matrix $\widetilde{B}_{\alpha }^{\ \alpha ^{\prime }}(u,\theta )$ of
parametric transforms can be computed by introducing the relations (\ref{4qe}%
), (\ref{4qef}) into (\ref{4ger1t}),
\begin{equation}
\widetilde{B}_{\alpha }^{\ \alpha ^{\prime }}=\widetilde{A}_{\alpha }^{\
\underline{\alpha }}\ A_{\underline{\alpha }}^{\ \alpha ^{\prime }}
\label{4mgt}
\end{equation}%
where $A_{\underline{\alpha }}^{\ \alpha ^{\prime }}$ is inverse to $%
A_{\alpha ^{\prime }}^{\ \underline{\alpha }}.$

The second parametric method \cite{4geroch2} was similarly developed which
yields a family of new exact solutions involving two arbitrary functions on
one variables, beginning with any two commuting Killing fields for which a
certain pair of constants vanish (for instance, the exterior field of a
rotating star). By successive iterating such parametric transforms, one
generates a class of exact solutions characterized by an infinite number of
parameters and involving arbitrary functions. For simplicity, in this work
we shall consider only a nonholonomic version of the first parametric method.


\begin{thebibliography}{99}
\bibitem{4ham1} R. S. Hamilton, Three Manifolds of Positive Ricci Curvature,
J. Diff.\ Geom. \textbf{17} (1982) 255--306

\bibitem{4ham2} R.\ S. Hamilton, The Formation of Singularities in the Ricci
Flow, in: \textit{\ Surveys in Differential Geometry}, Vol. 2 (International
Press, 1995), pp. 7--136

\bibitem{4per1} G. Perelman, The Entropy Formula for the Ricci Flow and its
Geometric Applications, math.DG/ 0211159

\bibitem{4caozhu} H. -D. Cao and X. -P. Zhu, Hamilton--Perelman's Proof of
the Poincare Conjecutre and the Geometrization Conjecture, Asian J. Math.,
\textbf{10 } (2006) 165--495, math.DG/0612069

\bibitem{4kleiner} B. Kleiner and J. Lott, Notes on Perelman's Papers,
math.DG/0605667

\bibitem{4rbook} J.\ W. Morgan and G. Tian, Ricci Flow and the Poincare
Conjecture, math.DG/0607607

\bibitem{4ni} M. Nitta, Conformal Sigma Models with Anomalous Dimensions and
Ricci Solitons, Mod. Phys. Lett. A \textbf{\ 20 } (2005) 577-584

\bibitem{4cv} S. A. Carstea and M. Visinescu, Special Solutions for Ricci Flow
Equation in 2D Using the Linearization Approach, Mod. Phys. Lett. A \textbf{%
\ 20 } (2005) 2993-3002

\bibitem{4gk} J. Gegenberg and G. Kunstatter, Ricci Flow of 3-D Manifolds
with One Killing Vector, J. Math. Phys. \textbf{47 } (2006) 032304

\bibitem{4osw} T. Oliynyk, V. Suneeta and E. Woolgar, A Gradient Flow for
Worldsheet Nonlinear Sigma Models, Nucl. Phys. B \textbf{\ 739 } (2006)
441-458

\bibitem{4hw} M. Headrick and T. Wiseman, Ricci Flow and Black Holes, Class.
Quantum. Grav. \textbf{\ 23 } (2006) 6683-6707

\bibitem{4bop} I. Bakas, D. Orlando and P. M. Petropoulos, Ricci Flows and
Expansion in Axion--Dilaton Cosmology, J. High Energy Phys. \textbf{01 }
(2007) 040

\bibitem{4nhrf01} S. Vacaru, Nonholonomic Ricci Flows:\ I.\ Riemann Metrics
and La\-grange--Finsler Geometry, math.DG/ 0612162

\bibitem{4nhrf02} S. Vacaru, Nonholonomic Ricci Flows: II. Evolution
Equations and Dynamics, J. Math. Phys. \textbf{ 49 } (2008) 043504

\bibitem{4nhrf03} S. Vacaru, Nonholonomic Ricci Flows: III. Curve Flows and
Solitonic Hierarchies, math.DG/ 0704.2062

\bibitem{4ma} R. Miron and M. Anastasiei, \textit{\ The Geometry of Lagrange
Spaces Theory and Applications } (Kluwer, 1994)

\bibitem{4bejf} A. Bejancu and H. R. Farran,\textit{\ Foliations and
Geometric Structures} (Springer, 2005)

\bibitem{4vesnc} S. Vacaru, Exact Solutions with Noncommutative Symmetries
in Einstein and Gauge Gravity, J. Math. Phys. \textbf{\ 46 } (2005) 042503

\bibitem{4vsgg} \textit{\ Clifford and Riemann- Finsler Structures in
Geometric Mechanics and Gravity}, Selected Works, by S. Vacaru, P.
Stavrinos, E. Gaburov and D. Gon\c{t}a. Differential Geometry -- Dinamical
Systems, Monograph 7 (Geometry Balkan Press, 2006);
www.mathem.pub.ro/dgds/mono/va-t.pdf and gr-qc/0508023

\bibitem{4vhep} S. Vacaru, Anholonomic Soliton-Dilaton and Black Hole
Solutions in General Relativity, JHEP, \textbf{04} (2001) 009

\bibitem{4vt} S. Vacaru and O. Tintareanu-Mircea, Anholonomic Frames,
Generalized Killing Equations, and Anisotropic Taub NUT Spinning Spaces,
Nucl. Phys. B \textbf{626} (2002) 239--264

\bibitem{4vp} S. Vacaru and F. C. Popa, Dirac Spinor Waves and Solitons in
Anisotropic Taub--NUT Spaces, Class. Quant. Grav. \textbf{18} (2001)
4921--4938

\bibitem{4vs} S. Vacaru and D. Singleton, Ellipsoidal, Cylindrical, Bipolar
and Toroidal Wormholes in 5D Gravity, J. Math. Phys. \textbf{43} (2002)
2486--2504

\bibitem{4entrnf} S. Vacaru, The Entropy of Lagrange--Finsler Spaces and
Ricci Flows, Rep. Math. Phys. \textbf{63} (2009) 95--110;   math.DG/0701621

\bibitem{4vrf} S.\ Vacaru, Ricci Flows and Solitonic pp--Waves, Int. J. Mod.
Phys. \textbf{A 21} (2006) 4899--4912

\bibitem{4vv1} S.\ Vacaru and M. Visinescu, Nonholonomic Ricci Flows and
Running Cosmological Constant:\ I.\ 4D Taub--NUT Metrics, Int. J. Mod. Phys.
\textbf{A 22} (2007) 1135--1159

\bibitem{4vv2} S. Vacaru and M. Visinescu, Nonholonomic Ricci Flows and
Running Cosmological Constant:\  3D Taub--NUT Metrics,
Romanian Reports in Physics \textbf{  60 } (2008) 218-238;\ gr--qc/ 0609086

\bibitem{4nhrf05} S. Vacaru, Nonholonomic Ricci Flows: V. Parametric
Deformations of Solitonic pp--Waves and Schwarzschild Solutions, math-ph/
0705.0729

\bibitem{4kramer} D. Kramer, H. Stephani, E. Herdlt and M. A. H. MacCallum,
Exact Solutions of Einstein's Field Equations (Cambridge University Press,
1980); 2d edition (2003)

\bibitem{4bic} J. Bicak, Selected Solutions of Einstein's Field Equations:
Their Role in General Relativity and Astrophysics, in:\ Lect. Notes. Phys.
\textbf{540} (2000), pp. 1--126

\bibitem{4geroch1} R. Geroch, A Method for Generating Solutions of
Einstein's Equation, J. Math. Phys. \textbf{12} (1971) 918--925

\bibitem{4geroch2} R. Geroch, A Method for Generating New Solutions of
Einstein's Equation. II, J. Math. Phys. \textbf{13} (1972) 394--404

\bibitem{4vpnhf} S. Vacaru, Parametric Nonholonomic Frame Transforms and
Exact Solutions in Gravity,
Int. J. Geom. Methods. Mod. Phys. (IJGMMP) \textbf{ 4 } (2007) 1285-1334;\
gr-qc/ 0704.3986

\bibitem{4esv} F. Etayo, R. Santamar\'{\i}a and S. Vacaru, Lagrange--Fedosov
Nonholonomic Manifolds, J. Math. Phys. \textbf{46} (2005) 032901

\bibitem{4string1} P. Deligne, P. Etingof, D. S. Freed et all (eds.),
\textit{\ Quanum Fields and Strings: A Course for Mathematicians}, Vols 1
and 2, Institute for Adavanced Study (American Mathematical Society, 1994)

\bibitem{4string2} J. Polchinski, \textit{String Theory,} Vols 1 \& 2
(Cambrdge Univ. Press, 1998)
\end{thebibliography}
\end{document}